\documentstyle[amssymb,aps,epsf]{revtex}

\begin{document}
\draft
\title{{\bf Gaussian Kinetic Model for Granular Gases}}
\author{James W. Dufty and Aparna Baskaran}
\address{Department of Physics, University of Florida, Gainesville, Florida}
\author{Lorena Zogaib}
\address{Departamento de Matem\'{a}ticas,\\ Instituto Tecnol\'{o}gico
Aut\'{o}nomo de M\'{e}xico\\
M\'{e}xico D. F., \ M\'{e}xico}
\date{December 29, 2003}
\maketitle

\begin{abstract}
A kinetic model for the Boltzmann equation is proposed and
explored as a practical means to investigate the properties of a
dilute granular gas. It is shown that all spatially homogeneous
initial distributions approach a universal "homogeneous cooling
solution" after a few collisions. The homogeneous cooling solution
(HCS) is studied in some detail and the exact solution is compared
with known results for the hard sphere Boltzmann equation. It is
shown that all qualitative features of the HCS, including the
nature of over population at large velocities, are reproduced
semi-quantitatively by the kinetic model. It is also shown that
all the transport coefficients are in excellent agreement with
those from the Boltzmann equation. Also, the model is specialized 
to one having a velocity independent collision frequency and the
resulting HCS and transport coefficients are compared to known
results for the Maxwell Model. The potential of the model for the
study of more complex spatially inhomogeneous states is discussed.
\end{abstract}

\pacs{PACS Numbers: 45.70.-n, 05.20.Dd, 51.10.+y}

\section{Introduction}

Many features of granular gases are captured by an idealized
system of smooth, hard spheres with inelastic collisions
\cite{brey3,ernst00rev}. During the past decade considerable
attention has been given to this simple system to understand
better the mechanisms behind observed qualitative differences
between real gases and those comprised of grains. Among the most
productive theoretical tools for analysis used is that of kinetic
theory \cite{vNyE01},
and more specifically at low density, the Boltzmann equation \cite%
{GyS95,dufty}. In recent years important conceptual issues, such
as the applicability of fluid dynamical equations, have been
clarified and quantitative methods have been developed for
accurate predictions over a wide range of experimental conditions.
It is fair to say that the Boltzmann kinetic theory is now the
primary theoretical method for a quantitative description of
granular gases.

While there are accurate and efficient numerical algorithms for
solving the Boltzmann equation \cite{bird}, analytic
approximations are more limited and exact solutions non-existent.
Such analytic results are of considerable interest because they
complement numerical solutions with a more penetrating explication
of the dominant mechanisms involved in a specific application. The
mathematical complexity of the Boltzmann collision operator is the
limiting factor in making progress, so simpler ''kinetic models''
have been proposed \cite{Cergignani75}. This approach has been
used with great success for real gases with elastic collisions
where several exact solutions far from equilibrium have been
obtained and shown to be in semi-quantitative
agreement with the numerical \ simulations of the \ Boltzmann equation \cite%
{Dufty90,santosbook}. Recent applications of kinetic models to granular
gases have yielded similar interesting exact results \cite%
{brey00model,brey97shear,tij01}. The collision operator for a kinetic model
is constrained to preserve the most important exact properties of the
Boltzmann collision operator, such as a special homogeneous solution and the
macroscopic balance equations for mass, momentum, and energy. Otherwise, the
model is chosen for simplicity and tractability. The objective here is to
recall one of the first kinetic models proposed for granular gases, the BMD model
 \cite{BMyD96}, and to generalize it for both a qualitative and a quantitative
representation of the Boltzmann equation. It will be referred to as the
Gaussian model for reasons that will become apparent.

In the following sections the Gaussian model is defined and
applied to the simplest cases of homogeneous states and weak
spatial perturbations of those states. The motivation for this
work is to provide a tool for a subsequent more detailed study of
spatially inhomogeneous states. For example, recent results
suggest that the spectrum of the linearized collision operator for
a realistic kinetic model could shed important light on the
validity conditions for a hydrodynamic description
\cite{DyB03,BDyR03}. Also, a more practical means to describe
boundary value problems is desired for a more faithful comparison
with experiments. Attention is focused here on spatially
homogeneous states for an isolated system and on transport
coefficients for small spatial perturbations, as a means to
compare and contrast models. A more detailed application to
inhomogeneous states and shear flow will be presented elsewhere.

There is extensive current interest in related Maxwell models.
Several exact results have been obtained recently for the
homogeneous state of an isolated system using these models
\cite{BNyK03}. In some mathematical respects the Maxwell models
are closer to the hard sphere Boltzmann collision operator than
the model studied here. However, its predictions (e.g.,
homogeneous cooling state distribution, transport coefficients)
are quite different from those of the Boltzmann equation as
recounted below. In contrast, the Gaussian\ model is structurally
simpler but with the capacity to give a better representation of
known results for the Boltzmann equation.

The basic results known for the hard sphere Boltzmann equation are
summarized in the next section. The ideas of kinetic modeling and
some existing models are reviewed briefly in section 3, and the
Gaussian model is defined in section 4. It is an extension and
synthesis of two earlier models, the ellipsoidal statistical (ES)
model \cite{ES66} introduced for gases with elastic collisions to
yield the correct Prandtl number, and the BMD model noted above
for inelastic collisions. The new model is constructed to describe
both elastic and inelastic collisions, retain a realistic velocity
dependent collision frequency, and to yield the correct Prandtl
number for accurate transport coefficients.

Also in Section 4, the exact solution to the Gaussian model is
obtained for an arbitrary initial homogeneous state. It is shown
that this class of solutions approaches a universal HCS solution
on a time scale of the order of several collisions. Hence the HCS
is the special state for homogeneous granular gases analogous to
the Maxwellian for  normal gases. The properties of the HCS
distribution are shown to be similar to those for the Boltzmann
equation at both small and large velocities. In particular similar
exponential decay occurs at large velocities. The special case of
a velocity independent collision frequency is studied as well. In
that case the HCS is the same as that for the BMD model. The
Maxwell models also have a velocity independent collision
frequency. In this special case Gaussian model HCS has properties
quite similar to those of the Maxwell models, including algebraic
decay for large velocities.

The Chapman-Enskog solution to the Boltzmann equation for small
spatial variations around the local HCS applies for the models as
well. To Navier-Stokes order this solution is characterized by
three transport coefficients. These transport coefficients are
compared for the various models and the Boltzmann equation in
section 5. The Gaussian model with velocity dependent collision
frequency provides an excellent representation of the hard sphere
Boltzmann results, over a wide range of inelasticity.

The HCS is the reference state for linear hydrodynamics.
Consequently, knowledge of the exact form of this distribution
function from the kinetic models provides some additional insight
there as well. This is illustrated
in Section 6 where the velocity dependence of the hydrodynamic modes \cite%
{BDyR03,DyB03} is calculated. The relationship of these modes to the fluxes
in Green-Kubo expressions for transport coefficients is also noted. Finally,
the results presented here are summarized in the last section and some other
interesting applications of the kinetic model are suggested.

\section{Hard sphere Boltzmann equation}

\label{sec2}

The system considered is composed of $N$ smooth hard spheres of diameter $%
\sigma $ in a large volume $V$. If the density is sufficiently small, $%
N\sigma ^{3}/V<<1$, the one-particle distribution function, $f({\bf r},{\bf v%
},t)$, for the number of particles with position ${\bf r}$ and velocity $%
{\bf v}$ at time $t$ is determined from the Boltzmann equation \cite%
{GyS95,dufty}
\begin{equation}
\left( \frac{\partial }{\partial t}+{\bf v}\cdot \nabla \right) f({\bf r},%
{\bf v},t)=J\left( {\bf r},{\bf v}|f(t)\right) .  \label{2.1}
\end{equation}%
The Boltzmann collision operator, $J$, has the form
\begin{equation}
J\left( {\bf r},{\bf v}|f(t)\right) \equiv -\nu ({\bf r},t,v)f({\bf r},{\bf v%
},t)+\int d{\bf v}_{1}\int d\widehat{{\bf \sigma }}\,K(\widehat{\sigma }%
\cdot {\bf g}^{\prime })\alpha ^{-1}f({\bf r},{\bf v}^{\prime },t,)f({\bf r},%
{\bf v}_{1}^{\prime },t).  \label{2.2}
\end{equation}%
The first term on the right side represents the loss of particles with
velocity ${\bf v}$ at a rate due to the collision frequency $\nu ({\bf r}%
,v,t)$
\begin{equation}
\nu ({\bf r},v,t)=\int d{\bf v}_{1}\int d\widehat{{\bf \sigma }}\,K(\widehat{%
\sigma }\cdot {\bf g})f({\bf r},{\bf v}_{1},t).  \label{2.2b}
\end{equation}%
The second term of (\ref{2.2}) represents the gain of particles with
velocity ${\bf v}$, where $\left\{ {\bf v}^{\prime },{\bf v}_{1}^{\prime
}\right\} $ are the ''restituting'' velocities that lead to $\left\{ {\bf v},%
{\bf v}_{1}\right\} $ following a smooth, inelastic hard sphere collision
\begin{equation}
{\bf v}^{\prime }={\bf v}-\frac{1}{2}(1+\alpha ^{-1})(\widehat{{\bf \sigma }}%
\cdot {\bf g})\widehat{{\bf \sigma }},\hspace{0.3in}{\bf v}_{1}^{\prime }\,=%
{\bf v}_{1}+\frac{1}{2}(1+\alpha ^{-1})(\widehat{{\bf \sigma }}\cdot {\bf g})%
\widehat{{\bf \sigma }}.  \label{2.3}
\end{equation}%
Here, $\widehat{\sigma }$ is a unit vector along the line of their centers,
and ${\bf g}={\bf v}-{\bf v}_{1}$. The parameter $\alpha $ is the
coefficient of restitution, $0<\alpha \leq 1$, describing the fractional
change in the normal component of the relative velocity ($\widehat{{\bf %
\sigma }}\cdot {\bf g}=-\alpha \widehat{{\bf \sigma }}\cdot {\bf
g}^{\prime } $) and hence the inelasticity of collisions
($\alpha=1$ corresponds to elastic collisions. The kernel
$K(\widehat{\sigma }\cdot {\bf g}) $ is proportional to the flux
of particles times the differential cross section and is given by
\begin{equation}
K(\widehat{\sigma }\cdot {\bf g})=\sigma ^{2}\Theta (\widehat{\sigma }\cdot
{\bf g})(\widehat{{\bf \sigma }}\cdot {\bf g}),  \label{2.2a}
\end{equation}%
where $\Theta $ is the Heaviside step function .

The most important properties of the collision operator are those that
result from the microscopic balance equations for mass, momentum, and energy
in a two particle collision. For the collision rules (\ref{2.3}) it follows
directly that $J$ has the following exact properties
\begin{equation}
\int d{\bf v}\left(
\begin{array}{c}
1 \\
{\bf v} \\
\frac{1}{2}m\left( {\bf v}-{\bf u}\right) ^{2}%
\end{array}
\right) J\left( {\bf r},{\bf v}|f(t)\right) =\left(
\begin{array}{c}
0 \\
{\bf 0} \\
-\frac{3}{2}nT\zeta%
\end{array}
\right) ,  \label{2.4}
\end{equation}
where $m$ is the mass, $n$ is the density$,T$ is the temperature$,$ and $%
{\bf u}$ is the macroscopic flow velocity
\begin{equation}
\left(
\begin{array}{c}
n({\bf r},t) \\
n({\bf r},t){\bf u}({\bf r},t) \\
\frac{3}{2}n({\bf r},t)T({\bf r},t)%
\end{array}
\right) =\int d{\bf v}\left(
\begin{array}{c}
1 \\
{\bf v} \\
\frac{1}{2}m\left( {\bf v}-{\bf u}\right) ^{2}%
\end{array}
\right) f({\bf r},{\bf v,}t).  \label{2.4a}
\end{equation}
The two zeros on the right side of (\ref{2.4}) correspond to conservation of
mass and momentum. The last term results from non-conservation of energy and
implies the cooling equation for homogeneous states
\begin{equation}
T^{-1}\partial _{t}T=-\zeta,  \label{2.4c}
\end{equation}
where $\zeta $ is the ''cooling rate''
\begin{equation}
\zeta =\left( 1-\alpha ^{2}\right) \frac{m\pi \sigma ^{2}}{24nT}\int \,d{\bf %
v}\,\int \,d{\bf v}_{1}\,g^{3}\ f({\bf r},{\bf v},t,)f({\bf r},{\bf v}%
_{1},t).  \label{2.5}
\end{equation}

It is easy to verify from (\ref{2.4}) that there is no spatially
homogeneous steady state for the isolated system, in contrast to
gases with elastic collisions. Instead, there exists a special
solution, the homogeneous cooling solution (HCS), which is assumed
to be approached in a few collision times by all homogeneous
initial conditions. It has a scaling property such that the
dependence on time occurs only through the temperature
\begin{equation}
f_{hcs}({\bf v,}t)=nv_{0}^{-3}(t)\phi \left( v^{\ast }\right) ,\hspace{0.25in%
}v^{\ast }=v/v_{0}(t),\hspace{0.25in}v_{0}(t)=\sqrt{2T(t)/m}.  \label{2.6}
\end{equation}%
In the following $v_{0}(t)$ will be referred to as the thermal velocity in
analogy to a gas with elastic collisions. Substitution of (\ref{2.6}) into
the Boltzmann equation leads to a time independent equation for $\phi \left(
v^{\ast }\right) $ that must be solved self-consistently with the
determination of $\zeta $ from (\ref{2.5})
\begin{equation}
\frac{1}{2}\zeta ^{\ast }\left( 3+v^{\ast }\frac{\partial }{\partial v^{\ast
}}\right) \phi \left( v^{\ast }\right) =J^{\ast }\left( v^{\ast }|\phi
\right) ,\hspace{0.25in}\zeta ^{\ast }\equiv \frac{\zeta }{\nu _{0}},\hspace{%
0.25in}J^{\ast }=\frac{J}{\nu _{0}},  \label{2.6a}
\end{equation}%
where $\nu _{0}$ is an average collision frequency
\begin{equation}
\nu _{0}(t)=\frac{16}{5}n\sigma ^{2}\sqrt{\frac{\pi }{2}}v_{0}(t).
\label{2.6b}
\end{equation}%
Because of the scaling property of $f_{hcs}$ both $\zeta^{\ast}$
and $J^{\ast}$ are independent of time. This problem has been
studied in detail in recent years and only a few
results will be quoted here. For $v^{\ast }\leq 2$ the solution to (\ref%
{2.6a}) can be expanded in Sonine polynomials about a Maxwellian with the
result \cite{vNyE98}

\begin{equation}
\phi \left( v^{\ast }\right) \rightarrow \pi ^{-3/2}e^{-v^{\ast 2}}\left[ 1+%
\frac{c(\alpha )}{4}\left( v^{\ast 4}-5v^{\ast 2}+\frac{15}{4}\right) \right]
,  \label{2.7}
\end{equation}
with
\begin{equation}
c(\alpha )\text{ }\rightarrow \text{ }c_{B}(\alpha )=\frac{32\left( 1-\alpha
\right) \left( 1-2\alpha ^{2}\right) }{81-17\alpha +30\alpha ^{2}\left(
1-\alpha \right) }.  \label{2.9}
\end{equation}
The subscript $B$ has been included on $c_{B}(\alpha )$ here to distinguish
the value determined by the hard sphere Boltzmann equation from that for the
models introduced below. The accuracy \ of the representation (13) at small $%
v$ to within a few percent for all $\alpha $ has been confirmed by Monte
Carlo simulation \cite{brey96hcs,santos00a}. The cooling rate calculated
from (\ref{2.7}) is \cite{BDKyS98}
\begin{equation}
\zeta ^{\ast }=\frac{5}{12}\left( 1-\alpha ^{2}\right) \left( 1+\frac{3}{32}%
c_{B}(\alpha )\right).  \label{2.10}
\end{equation}
For asymptotically large velocities $\phi \left( v^{\ast }\right) $ has a
qualitatively different behavior \cite{vNyE98}
\begin{equation}
\phi \left( v^{\ast }\right) \rightarrow Ae^{-\frac{2\beta_{B}}{\zeta^{\ast}}v^{\ast }},\hspace{%
0.25in}\beta_{B}(\alpha )=\frac{5\sqrt{2\pi }}{16}. \label{2.12}
\end{equation}
The constant $\beta_{B}$ arises in the large velocity limit of the
collision frequency,
\begin{equation}
\nu^{\ast}(v^{\ast})\equiv \frac{\nu(v,t)}{\nu_{o}(t)}\rightarrow
\beta_{B}v^{\ast}. \label{2.12c}
\end{equation}
Thus the origin of the exponential decay is the asymptotic
velocity dependence of the collision frequency. The over
population in the tail of the distribution, relative to the
Gaussian at small velocities, also has been confirmed for $v^{\ast
}\geq2$ for all $\alpha $ by Monte Carlo simulation
\cite{brey99,santos00a}. These and the transport properties of
Section 5 are the primary results known for the Boltzmann equation
with inelastic collisions. They are the main features captured by
the kinetic model proposed here.

\section{An Overview of Kinetic Models}
It is remarkable that over a century after Boltzmann wrote his
kinetic equation for a low density gas, the content of that
equation remains masked by its complexity. Certainly, a great deal
is known about solutions near the equilibrium state but the
mechanisms controlling nonlinear transport far from equilibrium
are still poorly understood. Significant progress has been made in
the past twenty years with the development of Direct Simulation
Monte Carlo methods (DSMC) by Bird \cite{bird}. This numerical
tool is exceptionally powerful and provides access now to a wide
range of nonequlibrium states for both elastic and inelastic
collisions. For more detailed analytical insight, kinetic models
have provided a parallel powerful tool in rarefied gas dynamics.
The objective of this section is to give a brief summary of the
concept of kinetic models and their extension to inelastic
collisions. Although the discussion is limited to the Boltzmann
equation, it is noted that the same ideas have been applied as
well to its dense fluid generalization, the Enskog kinetic
equation, for both fluids and solids\cite{dufty98,Lutsko97}.

\subsection{Maxwell Model}

The results quoted in the previous section for the HCS are
accurate but not exact. To obtain a more penetrating investigation
of this and other solutions a simplified
version of the Boltzmann equation called the Maxwell model has been proposed %
\cite{BN00} whereby the kernel $K(\widehat{\sigma }\cdot {\bf g})$ is
replaced by a velocity independent kernel $K({\bf r},t)$.Then (\ref{2.2b})
implies that the collision frequency also is independent of the velocity, $%
\nu ({\bf r},t)=4\pi n({\bf r},t)K({\bf r},t)$. The resulting model for the
Boltzmann collision operator becomes \cite{santos03}
\begin{equation}
J\left( {\bf r},{\bf v}|f(t)\right) \rightarrow J_{M}\left( {\bf r},{\bf v}%
|f(t)\right) \equiv -\nu ({\bf r},t)\left[ f({\bf r},{\bf v},t)-\frac{1}{%
4\pi n({\bf r},t)}\int d{\bf v}_{1}\int d\widehat{{\bf \sigma }}\,\alpha
^{-1}f({\bf r},{\bf v}^{\prime },t,)f({\bf r},{\bf v}_{1}^{\prime },t)\right]
.  \label{2.13}
\end{equation}
The collision frequency $\nu ({\bf r},t)$ is a free parameter of
the model. Its dependence on space and time is due to a possible
functional dependence on $f({\bf r},{\bf v},t).$ To fix $\nu ({\bf
r},t)$ the cooling rate is calculated directly for the Maxwell
model with the result
\begin{equation}
\zeta _{M}=\frac{1}{6}\left( 1-\alpha ^{2}\right) \nu ({\bf r},t).
\label{2.14}
\end{equation}
The collision frequency is now chosen to assure that the cooling rate for
the model is the same as that for the hard sphere Boltzmann equation, $%
\zeta_{M}=\zeta$, given to good approximation by (\ref{2.10}). This requires
the choice
\begin{equation}
\nu ({\bf r},t)=\frac{5}{2}\left( 1+\frac{3}{32}c_{B}(\alpha )\right) \nu
_{0}({\bf r},t),  \label{2.15}
\end{equation}
where $\nu _{0}$ is given by (\ref{2.6b}). This completely fixes the Maxwell
model.

The approximate Maxwell form for the Boltzmann collision operator
does not represent any real kinetics due to scattering by a
potential. It is called a Maxwell model because the property of
$K$ being independent of the velocity follows for scattering by
Maxwell molecules interacting via an inverse fourth power law
potential. However, the model described here retains the collision
rules for inelastic hard spheres, (\ref{2.3}), and therefore is a
hybrid not corresponding to any potential. Still, it provides an
interesting and tractable model for which several exact results
have been obtained recently.

The HCS has been studied for this Maxwell \ model as well. For small
velocities $\phi \left( v^{\ast }\right) $ again has the form of (\ref{2.7})
except that the coefficient $c(\alpha )$ is replaced by \cite{santos03}
\begin{equation}
c(\alpha )\text{ }\rightarrow c_{M}(\alpha )=\frac{12\left( 1-\alpha \right)
^{2}}{5+3\alpha \left( 2-\alpha \right) }.  \label{2.16}
\end{equation}
This is significantly different from the small velocity dependence
of the hard sphere Boltzmann equation, suggesting that the Maxwell
model does not reproduce quantitatively the HCS solution for hard
spheres. Furthermore, the difference is even qualitative at larger
velocities. The exact asymptotic behavior from the Maxwell model
is
\begin{equation}
\phi \left( v^{\ast }\right) \rightarrow Av^{\ast -k\left( \alpha \right) }.
\label{2.17}
\end{equation}
Thus there is algebraic decay for the Maxwell model in contrast to the
exponential decay for hard spheres. The exponent $k\left( \alpha \right) $
is the solution to a transcendental equation \cite{BNyK02,ernst02}
\begin{equation}
1+\frac{1}{12}\left( 1-\alpha ^{2}\right) \left( 3-k\right) =\left( \frac{%
1+\alpha }{2}\right) ^{k-3}\frac{\Gamma \left( \frac{k-2}{2}\right) }{%
2\Gamma \left( \frac{k}{2}\right) }+\int_{0}^{1}dx\left( 1-\left( 1-\frac{1}{%
4}\left( 1-\alpha \right) ^{2}\right) x^{2}\right) ^{\left( k-3\right) /2}.
\label{2.18}
\end{equation}
The behavior of $c_{M}\left(\alpha\right)$ and $k\left( \alpha \right) $ is
illustrated in the next section.

The transport coefficients associated with Navier-Stokes
hydrodynamics also have been calculated for the Maxwell
model\cite{santos03}. The agreement with those from the Boltzmann
equation for \ hard spheres is only qualitative (see Section 5).
While the Maxwell model allows interesting and exact solutions, it
does not appear to provide a reliable representation of the
Boltzmann equation for \ hard spheres and therefore the results
obtained from it must be interpreted with some care.

\subsection{Other Kinetic Models}

The Maxwell model, while simpler than the Boltzmann equation is
still quite complex and even for the HCS the exact distribution
function has been calculated only  in one dimension. Historically,
for normal gases, a number of simpler kinetic models have been
applied with great success. More recently, these models have been
extended to granular gases with a similar success in applications.
To explain them generically, it is useful to rewrite the Boltzmann
equation (\ref{2.1}) to make the effects of cooling explicit
\cite{BDyS99}
\begin{equation}
\left( \frac{\partial }{\partial t}+{\bf v}\cdot \nabla \right) f({\bf r},%
{\bf v},t)-\frac{1}{2}\zeta \nabla _{{\bf v}}\cdot \left( {\bf V}f\right)
=J^{\prime }[{\bf r},{\bf v}|f(t)],  \label{3.1}
\end{equation}
with
\begin{equation}
J^{\prime }\left( {\bf r},{\bf v}|f(t)\right) =J\left( {\bf r},{\bf v}%
|f(t)\right) -\frac{1}{2}\zeta \nabla _{{\bf v}}\cdot \left( {\bf V}f\right)
,  \label{3.2}
\end{equation}
where ${\bf V}={\bf v}-{\bf u}$ is the velocity relative to the average
flow. Then the condition (\ref{2.4}) becomes
\begin{equation}
\int d{\bf v}\left(
\begin{array}{c}
1 \\
{\bf v} \\
\frac{1}{2}mV^{2}%
\end{array}
\right) J^{\prime }\left( {\bf r},{\bf v}|f(t)\right) =\left(
\begin{array}{c}
0 \\
{\bf 0} \\
0%
\end{array}
\right) .  \label{3.3}
\end{equation}
In addition, there is a null space for $J^{\prime }\left( {\bf r},{\bf v}%
|f(t)\right) $%
\begin{equation}
J^{\prime }\left( {\bf r},{\bf v}|f_{0}(t)\right) =0.  \label{3.4}
\end{equation}
The conditions (\ref{3.3}) and (\ref{3.4}) are the same as those for the
conservation laws and the equilibrium state, respectively, for elastic
collisions. More generally, (\ref{3.4}) defines the HCS in agreement with (%
\ref{2.6a}). These two sets of conditions are necessary for the macroscopic
balance equations (precursors to hydrodynamics) and the ''universal''
homogeneous state $f_{0}$. The basic idea of kinetic models is to replace
the actual Boltzmann collision operator by a simpler structure, while
preserving the properties (\ref{3.3}) and (\ref{3.4}).

There are many ways that a kinetic model can \ be constructed with these
constraints. Perhaps the simplest are the BGK model(s)\cite{Cergignani75}
\begin{equation}
J^{\prime }\left( {\bf r},{\bf v}|f_{0}(t)\right) \rightarrow -\nu \left(
{\bf r},t\right) \left( f\left( {\bf r},{\bf v},t\right) -f_{0}\left( {\bf r}%
,{\bf v},t\right) \right) .  \label{3.5}
\end{equation}
Clearly, (\ref{3.4}) is satisfied and the conditions (\ref{3.3}) are imposed
by requiring that the relevant moments of $f$ and $f_{0}$ should be the same
\begin{equation}
\int d{\bf v}\left(
\begin{array}{c}
1 \\
{\bf v} \\
\frac{1}{2}mV^{2}%
\end{array}
\right) \left( f\left( {\bf r},{\bf v},t\right) -f_{0}\left( {\bf r},{\bf v}%
,t\right) \right) =\left(
\begin{array}{c}
0 \\
{\bf 0} \\
0%
\end{array}
\right) .  \label{3.6}
\end{equation}
This implies that $f_{0}$ is a functional of $f$ so the apparent simplicity
of (\ref{3.5}) is misleading. For elastic collisions $f_{0}$ is taken to be
the local Maxwellian for consistency with the known equilibrium state. In
the case of inelastic collisions, it would seem appropriate to choose $f_{0}$
as the HCS distribution from the Boltzmann equation. However, since this is
not known it is more common to choose again $f_{0}$ as the local Maxwellian.
As described in the previous section, this is a reasonable first
approximation to the HCS if the velocities are not too large. However, it
precludes use of the kinetic model to study the HCS itself. The collision
frequency $\nu \left( {\bf r},t\right) $ is a free parameter of the model,
usually chosen to fit one of the transport coefficients. On dimensional
grounds $\nu \left( {\bf r},t\right) \propto n\left( {\bf r},t\right)
T^{1/2}\left( {\bf r},t\right) $ and therefore also a functional of $f$.

The Chapman-Enskog solution to the BGK kinetic equation for
inelastic collisions has been obtained to derive the associated
hydrodynamic equations to Navier-Stokes order (see Section 5)
\cite{BDKyS98}. The dependence of all transport coefficients on
the restitution coefficient $\alpha $ is in good semi-quantitative
agreement with that for the Boltzmann equation. However, the model
suffers from the same well-known \ problem for elastic collisions
of an incorrect Prandtl number $\eta C_{p}/\kappa $ where
$C_{p}=5k_{B}/2m$ is the specific heat per unit mass, $\eta$ is
the shear viscosity and $\kappa$ is the thermal conductivity.
Since the BGK model has only one parameter $\nu$ the absolute
value of either the shear viscosity or the thermal conductivity is
wrong by a factor of approximately 2/3. This can be corrected by
choosing for $f_{0}$ a more general Gaussian, with an additional
parameter leading to the ES model for elastic collisions
\cite{ES66} . BGK models of this type for granular gases have been
discussed recently by Astillero and Santos \cite{Astillero03}.

A related but different kinetic model attempts to represent more directly
the gain term of the Boltzmann collision operator \cite{BMyD96}. Equation (%
\ref{2.2}) is written as
\begin{equation}
J^{\prime }\left( {\bf r},{\bf v}|f(t)\right) \equiv -\nu ({\bf r,}%
V,t)\left( f({\bf r},{\bf v},t)-g({\bf r},{\bf V},t\mid f)\right) -\frac{1}{2%
}\zeta \nabla _{{\bf v}}\cdot \left( {\bf V}f\right) .  \label{3.7}
\end{equation}%
The gain functional $g({\bf r},{\bf V},t\mid f)$ is now chosen for
convenience and simplicity to \ define the model, but restricted by the
exact conditions (\ref{3.3}) and (\ref{3.4}).

In contrast to the BGK model, the condition (\ref{3.4}) now
provides an equation that determines a non-trivial HCS solution.
The simplest choice for $g({\bf r},{\bf V},t\mid f)$ is again a
Maxwellian, but with the
temperature modified to account for the extra term on the right side of (\ref%
{3.7}). For simplicity, applications of this kinetic model to date have also
chosen a velocity independent collision frequency. In the limit of elastic
collisions it reduces to the BGK model.

The HCS solution can be obtained exactly for this  model and, like
the Maxwell model, it has algebraic rather than exponential decay
at large velocities. The transport coefficients for this second
model are of comparable accuracy to those from the BGK model and
suffer from the same difficulty of an
incorrect Prandtl number. In the next section the kinetic model based on (%
\ref{3.7}) is generalized to include a velocity dependent
collision frequency and a Gaussian form for $g({\bf r},{\bf
V},t\mid f)$ that can accommodate the correct Prandtl number.

\section{Gaussian Kinetic Model}

The main objective of the present work is to propose a synthesis
of the BMD and ES models and to extend them to include a velocity
dependent  collision frequency. This will be referred to as the
Gaussian model. Like the Maxwell model, the Gaussian model admits
exact analysis in many interesting cases, but it is simpler and
captures more accurately the qualitative features of the Boltzmann
equation. In this section the model is defined and the initial
value  problem is solved exactly for spatially homogeneous states.
It is shown that all initial states rapidly  approach a universal
HCS. The HCS is then studied and compared with known results for
the Boltzmann equation for hard spheres. Finally, it is
specialized to the case of a velocity independent collision
frequency, for comparison with the HCS for the Maxwell model.

The model is defined by the choice of a Gaussian for $g({\bf r},{\bf v}%
,t\mid f)$ in (\ref{3.7}) \cite{hist1}
\begin{equation}
\left( \frac{\partial }{\partial t}+{\bf v}\cdot \nabla \right) f({\bf r},%
{\bf v},t)=-\nu ({\bf r},V,t)\left( f({\bf r},{\bf v},t)-g({\bf r},{\bf V}%
,t\mid f)\right) ,  \label{4.0}
\end{equation}
\begin{equation}
g({\bf r},{\bf V},t\mid f)=A({\bf r},t)e^{-V_{i}B_{ij}^{-1}({\bf r},t)V_{j}}.
\label{4.1}
\end{equation}
The scalar function $A({\bf r},t)$ and symmetric tensor
$B_{ij}({\bf r},t)$ are determined in part by the conditions
(\ref{2.4a})
\begin{equation}
\int d{\bf v}\left(
\begin{array}{c}
1 \\
{\bf v} \\
\frac{1}{2}mV^{2}%
\end{array}
\right) \nu ({\bf r},{\bf V},t)g({\bf r},{\bf V},t\mid f)=\left(
\begin{array}{c}
M_{1} \\
{\bf M}_{2} \\
M_{3}-\frac{3}{2}nT\zeta%
\end{array}
\right) ,  \label{4.2}
\end{equation}
where $M_{i}$ are moments of the distribution function, weighted by the
collision frequency
\begin{equation}
\left(
\begin{array}{c}
M_{1} \\
{\bf M}_{2} \\
M_{3}%
\end{array}
\right) =\int d{\bf v}\left(
\begin{array}{c}
1 \\
{\bf v} \\
\frac{1}{2}mV^{2}%
\end{array}
\right) \nu ({\bf r},V,t)f({\bf r},{\bf v},t).  \label{4.3}
\end{equation}
The choice of a Gaussian is primarily for convenience and simplicity.
However, it can be understood also as the result from information theory to
determine a function when only the moments in (\ref{4.2}) are specified (see
Appendix A). \ It follows directly that

\begin{equation}
{\bf M}_{2}({\bf r},t)=M_{1}({\bf r},t){\bf u(}r,t).  \label{4.5}
\end{equation}
For the special case of constant collision frequency \ these become
\begin{equation}
A\rightarrow n\left( \det \pi B\right) ^{-1/2};\hspace{0.25in}{\bf M}%
_{2}\rightarrow M_{1}{\bf u};\hspace{0.25in}\frac{1}{3}TrB\rightarrow \frac{T%
}{m}.  \label{4.6}
\end{equation}
so the coefficients of the Gaussian are related to the density, temperature,
and flow velocity. This also illustrates that the elements of $B$ are not
fully determined by the moment conditions (\ref{4.2}).

It is useful to divide the matrix $B_{ij}$ into a part proportional
to the unit matrix plus a traceless part
\begin{equation}
B_{ij}=B\delta _{ij}+\widetilde{B}_{ij},\hspace{0.25in}\widetilde{B}%
_{ij}=\left( B_{ij}-B\delta _{ij}\right) ,\hspace{0.25in}B=\frac{1}{3}B_{kk}.
\label{4.4}
\end{equation}
In the following a tilde above a \ matrix will be used to denote
its traceless part. The special case of (\ref{4.6}) shows that the
trace of $B_{ij}$ is proportional to the temperature and therefore
a linear functional of $f $. \ It is reasonable to choose the
remaining elements of $B_{ij}$ also to have a linear relationship
to $f$. Furthermore, it is required that this traceless part
should vanish at $f_{0}$, the solution to (\ref{3.4}), which
should be isotropic (to agree with the Boltzmann equation)
\begin{equation}
\widetilde{B}_{ij}=\int d{\bf v}\frac{\delta \widetilde{B}_{ij}}{\delta f({\bf v})}%
\mid _{f=f_{0}}\left( f-f_{0}\right) .  \label{4.7}
\end{equation}
Thus, the gain $g({\bf r},{\bf V},t\mid f)$ is anisotropic only
when evaluated for anisotropic states $f$.This is an implicit
definition since $f_{0}$ is a function of $B_{ij}$. However, the
form is such that $B_{ij}$ becomes diagonal when evaluated for
$f=f_{0}$, and hence so does $B_{ij}^{-1}$. This assures that
$f_{0}$, when it exists, is isotropic. The trace of $B_{ij}$ is a
scalar moment of degree 2 (including the weight factor $\nu $)
when $\widetilde{B}_{ij}=0$. Consequently, it is suggestive
to take $\widetilde{B}_{ij}$ as the traceless part of the moment of degree 2 of $%
f-f_{0}$. The final form for $B_{ij}$ is then
\begin{equation}
B_{ij}=B\delta _{ij}+\frac{y\left( \alpha \right) }{nm}\int d{\bf v}D_{ij}(%
{\bf V})\left( f-f_{0}\right) ,\hspace{0.25in}D_{ij}({\bf V})=m\left(
V_{i}V_{j}-\frac{1}{3}\delta _{ij}V^{2}\right),  \label{4.8}
\end{equation}
where $y\left( \alpha \right) $ is an undetermined dimensionless
quantity independent of the velocity. The conditions (\ref{4.2})
and (\ref{4.8}) completely determine the parameters $A,B_{ij}$.

It remains to choose the collision frequency $\nu ({\bf r},V,t)$
and the cooling rate $\zeta ({\bf r},t) $. In principle, these are
specific functionals of $f$ in the Boltzmann equation. Here, they
are taken to depend on $f$ only through the temperature and
density. The cooling rate is chosen to be the same as the
Boltzmann result (\ref{2.10})

\begin{equation}
\zeta ({\bf r},t)=\frac{5}{12}\left( 1-\alpha ^{2}\right) \left( 1+\frac{3}{%
32}c_{B}(\alpha )\right) \nu _{0}({\bf r},t),  \label{4.10}
\end{equation}
\begin{equation}
\nu _{0}({\bf r},t)=\frac{16}{5}n({\bf r},t)\sigma ^{2}\sqrt{\frac{\pi T(%
{\bf r},t)}{m}}.  \label{4.701}
\end{equation}

Similarly, guidance for the choice of the velocity dependent
collision frequency $\nu (\bf r,\bf V,t) $ is obtained from that
for the Boltzmann (see Eq(\ref{2.2b}))
\begin{equation}
\nu_{B}({\bf r},{\bf V},t)=\pi\sigma^{2}\int d{\bf V}_{1}f({\bf
r},{\bf V}_{1},t)|{\bf V}-{\bf V}_{1}| . \label{4.24}
\end{equation}
For small ${\bf V}$ this goes to a constant,
\begin{equation}
\nu_{B}\rightarrow \pi\sigma_{2}n({\bf
r},t)\overline{V},\hspace{0.25in}\overline{V}\equiv\frac{\int
d{\bf V}_{1} f V_{1}}{\int d{\bf V}_{1}f },
\end{equation}
while for large $\bf V$ it becomes linear in $V$.
\begin{equation}
\nu_{B}\rightarrow\pi\sigma{2}n({\bf r},t)V.
\end{equation}

 A representation of the complete
velocity dependence for the Gaussian model, preserving these
limiting forms is
obtained from (\ref{4.24}) using a Maxwellian for $f$%
\begin{equation}
\nu({\bf r},{\bf V},t)\equiv x(\alpha)\nu_{o}({\bf r},t)\nu
_{M}^{\ast }(v^{\ast }),\hspace{0.25in}\nu _{M}^{\ast }(v^{\ast
})=\frac{5\sqrt{2}}{16}\left[ e^{-v^{\ast
2}}+\left( 2v^{\ast }+v^{\ast -1}\right) \frac{\sqrt{\pi }}{2}%
\mathop{\rm erf}%
\left( v^{\ast }\right) \right] .  \label{4.26}
\end{equation}%
Here $x(\alpha)$ is a second undetermined dimensionless constant. The
particular choice for $x\left( \alpha \right) $ and the resulting
accuracy of the transport coefficients is discussed in the next
section. It is found that $x(\alpha )$ is a smooth function of
$\alpha $ of order unity. This form for $\nu ^{\ast }(v^{\ast })$
 has the correct large velocity dependence of (\ref{2.12}) but with
the coefficient differing by a factor of $x(\alpha)$ from the
Boltzmann equation.

At this point the Gaussian model has been specified in terms of
the two remaining constants $x(\alpha)$ and $y(\alpha)$. In the
next section it will be shown that the three transport
coefficients at Navier-Stokes order are functions of two
independent collision integrals. The constants $x(\alpha)$ and
$y(\alpha)$ are chosen to assure that these two collision
integrals are the same as those from the Boltzmann equation. This
leads to a coupled pair of equations ((\ref{5.15d}) and
(\ref{5.15c}) below) that are solved numerically. This completes
the definition of the Gaussian model.

\subsection{Spatially homogeneous states}

In the rest of this section, attention is restricted to spatially
homogeneous states. It is shown that for any arbitrary homogeneous
initial condition, the solution goes over to a universal HCS in a
few collision times. For these initial conditions  ${\bf u}({\bf
r},t)={\bf u}$ is constant and by a Galilean transformation it is
possible to choose ${\bf u=0}$. Also from the continuity equation
$n(t)=n$ is constant. The temperature obeys the cooling equation
(\ref{2.4c}) which is now written
\begin{equation}
T^{-1}\partial _{s}T=-\zeta ^{\ast },\hspace{0.25in}ds=\nu _{0}(t)dt,\hspace{%
0.25in}\zeta ^{\ast }=\frac{\zeta (t)}{\nu _{0}(t)}.  \label{4.71}
\end{equation}
The new time variable $s$ represents the average number of collisions in the
time $t$. It also follows from the definition of $\zeta (t)$ that $\zeta
^{\ast }$ is constant. The temperature therefore has a simple exponential
dependence on the collision number
\begin{equation}
T(s)=e^{-\zeta ^{\ast }s}T(0).  \label{4.72}
\end{equation}

Now consider a general homogeneous initial distribution and look for
solutions to the model kinetic equation in the dimensionless form
\begin{equation}
f\left( {\bf v},t\right) =nv_{0}^{-3}f^{\ast }\left( {\bf v}^{\ast
},s\right) ,\hspace{0.4cm}{\bf v}^{\ast }={\bf v}/v_{0}\left(
t\right) ,\hspace{0.4cm}\nu^{\ast}=\frac{\nu(v,t)}{\nu_{o}(t)}
.\label{4.73}
\end{equation}%
The dimensionless form for the model kinetic equation for
homogeneous states becomes
\begin{equation}
\left( \partial _{s}+\frac{1}{2}\zeta ^{\ast }\left( 3+{\bf v}^{\ast }\cdot
\nabla _{{\bf v}^{\ast }}\right) +\nu ^{\ast }\left( v^{\ast }\right)
\right) f^{\ast }\left( {\bf v}^{\ast },s\right) =\nu ^{\ast }\left( v^{\ast
}\right) g^{\ast }\left( {\bf v}^{\ast },s\mid f^{\ast }\right) ,
\label{4.75}
\end{equation}%
with
\[
g^{\ast }\left( {\bf v}^{\ast },s\mid f^{\ast }\right) =v_{0}^{3}g\left(
{\bf v},t\mid f\right) /n=A^{\ast }(s)e^{-v_{i}^{\ast }\left( B^{\ast
-1}\right) _{ij}(s)v_{j}^{\ast }}.
\]%
The moment ${\bf M}_{2}$ vanishes since ${\bf u}=0$. The remaining
dimensionless moments are
\begin{equation}
\left(
\begin{array}{c}
M_{1}^{\ast }\left( s\right) \\
M_{3}^{\ast }\left( s\right)%
\end{array}%
\right) =\left(
\begin{array}{c}
\frac{M_{1}(t)}{n\nu _{0}\left( t\right) } \\
\frac{2M_{3}(t)}{3nT\nu _{0}\left( t\right) }%
\end{array}%
\right) =\int d{\bf v}^{\ast }\left(
\begin{array}{c}
1 \\
\frac{2}{3}v^{\ast 2}%
\end{array}%
\right) \nu ^{\ast }(v^{\ast })f^{\ast }({\bf v}^{\ast },s).  \label{4.77}
\end{equation}%
The parameters $A^{\ast }(s)$ and $B_{ij}^{\ast }(s)$ are related to these
by
\begin{equation}
\int d{\bf v}^{\ast }\left(
\begin{array}{c}
1 \\
\frac{2}{3}v^{\ast 2}%
\end{array}%
\right) \nu ^{\ast }(v^{\ast })A^{\ast }(s)e^{-v_{i}^{\ast }\left( B^{\ast
-1}\right) _{ij}(s)v_{j}^{\ast }}=\left(
\begin{array}{c}
M_{1}^{\ast }\left( s\right) \\
M_{3}^{\ast }\left( s\right) -\zeta ^{\ast }%
\end{array}%
\right) ,  \label{4.77a}
\end{equation}%
\begin{equation}
B_{ij}^{\ast }=B^{\ast }\delta _{ij}+y\int d{\bf v}^{\ast }\left(
v_{i}^{\ast }v_{j}^{\ast }-\frac{1}{3}\delta _{ij}v^{\ast 2}\right) \left(
f^{\ast }-f_{0}^{\ast }\right) .  \label{4.77b}
\end{equation}

The formal solution to the kinetic equation is found in Appendix B with the
result
\begin{eqnarray}
f^{\ast }({\bf v}^{\ast },s) &=&e^{-\frac{3}{2}\zeta ^{\ast }s}K(v^{\ast
},s)f^{\ast }(e^{-\frac{1}{2}\zeta ^{\ast }s}{\bf v}^{\ast },0)  \nonumber \\
&&+\int_{0}^{s}ds^{\prime }e^{-\frac{3}{2}\zeta ^{\ast }s^{\prime
}}K(v^{\ast },s^{\prime })\nu ^{\ast }(e^{-\frac{1}{2}\zeta ^{\ast
}s^{\prime }}v^{\ast })g^{\ast }(e^{-\frac{1}{2}\zeta ^{\ast }s^{\prime }}%
{\bf v}^{\ast },s-s^{\prime }),  \label{4.78}
\end{eqnarray}
where $K(v^{\ast },s)$ represents the dynamics due to the loss term alone
\begin{equation}
K(v^{\ast },s)=\exp \left( -\int_{0}^{s}ds^{\prime \prime }\nu ^{\ast
}\left( e^{-\frac{1}{2}\zeta ^{\ast }s^{\prime \prime }}v^{\ast }\right)
\right) .  \label{4.79}
\end{equation}
The collision frequency is a monotonically increasing function of the
velocity so $\nu ^{\ast }\left( v^{\ast }\right) \geq \nu ^{\ast }\left(
0\right) $. This gives the inequality
\begin{equation}
K(v^{\ast },s)\leq e^{-\nu ^{\ast }\left( 0\right) s}.  \label{4.80}
\end{equation}
Since $\nu ^{\ast }\left( 0\right) $ is of order unity, the domain of
integration in\ (\ref{4.78}) is exponentially bounded for $s>1$ and for
large $s$ the integral becomes independent of $s$. The first term vanishes
exponentially fast and the $s$ independent HCS solution is obtained
\begin{equation}
f^{\ast }({\bf v}^{\ast },s)\rightarrow \phi (v^{\ast })=\int_{0}^{\infty
}ds^{\prime }e^{-\frac{3}{2}\zeta ^{\ast }s^{\prime }}K(v^{\ast },s^{\prime
})\nu ^{\ast }(e^{-\frac{1}{2}\zeta ^{\ast }s^{\prime }}v^{\ast })g^{\ast
}(e^{-\frac{1}{2}\zeta ^{\ast }s^{\prime }}v^{\ast },\infty )  \label{4.81}
\end{equation}
It is readily verified that $\phi (v^{\ast })$ is a stationary solution to (%
\ref{4.75}) and hence $\phi (v^{\ast })=f_{0}^{\ast }(v^{\ast })$ is the
unique HCS solution. It follows from the definition of $B$ that this
distribution becomes isotropic on the same time scale as (\ref{4.79}) even
if the initial distribution was not
\[
B_{ij}^{\ast }\left( s\right) \rightarrow \frac{1}{3}B_{kk}^{\ast }\left(
\infty \right) \delta _{ij}=B^{\ast }\delta _{ij}.
\]

For the class of homogeneous initial states considered, this
result shows that the HCS is the universal solution after several
collisions. Hence it is the special characteristic solution for
homogenous states analogous to
 the Maxwellian for elastic collisions. The result (\ref%
{4.81}) is stronger than the H-theorem for elastic collisions in
the sense that it implies the approach to the HCS is pointwise in
velocity space. It is interesting to observe that this analysis
does not require the explicit form for $\nu ^{\ast }(v^{\ast
})g^{\ast }(v^{\ast },s\mid \phi ^{\ast })$ and so it applies to
models with choices other than the Gaussian. In fact,
it applies to the Boltzmann equation itself although in that case (\ref{4.81}%
) is a more implicit functional relationship whose solution must be proved.
The consistency of the moment conditions (\ref{4.77}%
) and (\ref{4.77a}) is verified in Appendix C using the explicit form (\ref%
{4.81}). The functions $A^{\ast }=A^{\ast }\left( \infty \right) $ and $%
B^{\ast }=B_{kk}^{\ast }\left( \infty \right) /3$ are fixed by the fact that
$\widetilde{B}_{ij}^{\ast }=0$ and
\begin{equation}
\left(
\begin{array}{c}
1 \\
1%
\end{array}
\right) =\int d{\bf v}^{\ast }\left(
\begin{array}{c}
1 \\
\frac{2}{3}v^{\ast 2}%
\end{array}
\right) \phi (v^{\ast }).  \label{4.82}
\end{equation}
The explicit forms for these equations also are given in Appendix C.

\subsection{The Homogeneous Cooling State}

More explicit properties of the HCS are easily obtained.First, it
can be written in the more convenient form
\begin{equation}
\phi (v^{\ast })=\frac{2A^{\ast }}{\zeta ^{\ast }v^{\ast 3}}%
\int_{0}^{v^{\ast }}dxx^{2}\exp \left( -\frac{2}{\zeta ^{\ast }}%
\int_{x}^{v^{\ast }}\frac{dx^{\prime }}{x^{\prime }}\nu ^{\ast }\left(
x^{\prime }\right) \right) \nu ^{\ast }\left( x\right) e^{-B^{\ast -1}x^{2}}.
\label{4.27}
\end{equation}%
The coefficients $A^{\ast }$ and $B^{\ast }$ are determined from (\ref{4.82}%
). Both are smooth functions of $\alpha $ with the limiting values $A^{\ast
}=\pi ^{-3/2}$ and $B^{\ast }=1$ at $\alpha =1$. Practical fits for other
values of $\alpha $ in the range $0.4$ to $1$ are given by
\[
A^{\ast }=0.547-0.274\alpha -0.094\alpha ^{2},\hspace{0.25in}%
B^{\ast -1}=1.84-0.275\alpha -0.568\alpha ^{2}.
\]
Similarly, fits for $x(\alpha)$ and $y(\alpha)$ are found from
(\ref{5.15d}) and (\ref{5.15c}) to be
\[
x(\alpha)= 0.533 + 0.156\alpha -0.302\alpha^{2},
\]
\[
y(\alpha)=-0.906+2.666\alpha - 0.724\alpha^{2}.
\]
 For \ small $v^{\ast }$ the form of $\phi (v^{\ast
})$ is Gaussian
\begin{equation}
\phi (v^{\ast })\rightarrow C_{1}\exp -C_{2}v^{\ast 2},  \label{4.28}
\end{equation}%
with
\begin{equation}
C_{1}=\frac{A^{\ast }p}{3+p},\hspace{0.25in}C_{2}=\frac{\left( 3+p\right) }{%
\left( 5+p\right) }\left( B^{\ast -1}-\frac{3}{2\left( 3+p\right) \nu ^{\ast
}\left( 0\right) }\frac{d^{2}\nu ^{\ast }}{dv^{\ast 2}}\mid _{v^{\ast
}=0}\right),   \label{4.28a}
\end{equation}%
\begin{equation}
p=\frac{2\nu ^{\ast }(v^{\ast }=0)}{\zeta ^{\ast }}.  \label{4.28b}
\end{equation}%

The coefficient $c_{G}\left( \alpha \right) $ in the polynomial expansion\ (%
\ref{2.7}) for small $v^{\ast }$ is shown for comparison with the
corresponding Boltzmann hard sphere result in Figure 1. A
practical fit for $c_{G}$ in the range of $\alpha $ mentioned
above is given by
\[
c_{G}\left( \alpha \right) =0.247+0.865\alpha -2.907\alpha
^{2}+1.793\alpha ^{3}.
\]%
The model is seen to \ reproduce quite well the Boltzmann results for $%
\alpha \geq 0.8$ and has the same qualitative behavior for smaller $\alpha $%
. As indicated in Figure 1, the corresponding results for both the Maxwell \
model and the Gaussian model with velocity independent collision frequency
(to be discussed in the following section) are always positive and much
larger. This is the first of several observations showing an improvement of
the model due to the velocity dependent collision frequency. Figures 2
shows the exact distribution function reduced by the Maxwellian at $\alpha
=0.9$ . Also shown are the results from the polynomial expansion using $%
c_{G}\left( \alpha \right) $ and using $c_{B}(\alpha )$ for the
Boltzmann equation. It is seen that the polynomial expansion
follows the exact HCS closely and is close to the expansion using
$c_{B}(\alpha ).$ The polynomial expansion using $c_{B}(\alpha )$
is very close to the actual distribution obtained by Monte Carlo
simulation of the Boltzmann equation for $v^{\ast }<2 $. Therefore
the model gives a good representation of the Boltzmann
distribution for $v^{\ast }<2$. This is found to be true over the
whole range of $\alpha $.

The large $v^{\ast }$ dependence can be obtained as follows. First, rewrite (%
\ref{4.81}) as
\begin{equation}
\phi (v^{\ast })=\frac{I(v^{\ast },c)}{v^{\ast 3}}\exp \left( -\frac{2}{%
\zeta ^{\ast }}\int_{c}^{v^{\ast }}\frac{dx^{\prime }}{x^{\prime }}\nu
^{\ast }\left( x^{\prime }\right) \right) ,  \label{4.30}
\end{equation}%
\begin{eqnarray}
I(v^{\ast },c) &=&\frac{2A^{\ast }}{\zeta ^{\ast }}\int_{0}^{v^{\ast
}}dxx^{2}\exp \left( -\frac{2}{\zeta ^{\ast }}\int_{x}^{c}\frac{dx^{\prime }%
}{x^{\prime }}\nu ^{\ast }\left( x^{\prime }\right) \right) \nu ^{\ast
}\left( x\right) e^{-B^{\ast -1}x^{2}}  \nonumber \\
&=&I(\infty ,c)-\frac{2A^{\ast }}{\zeta ^{\ast }}\int_{v^{\ast }}^{\infty
}dxx^{2}\exp \left( -\frac{2}{\zeta ^{\ast }}\int_{x}^{c}\frac{dx^{\prime }}{%
x^{\prime }}\nu ^{\ast }\left( x^{\prime }\right) \right) \nu ^{\ast }\left(
x\right) e^{-B^{\ast -1}x^{2}}.  \label{4.30a}
\end{eqnarray}%
This form holds for any $c\leq v^{\ast }$. The integral satisfies
the bound
\begin{equation}
I(\infty,c)-I(v^{\ast } ,c)<I_{0}c^{2}\nu ^{\ast }\left( c\right)
e^{-B^{\ast -1}c^{2}},  \label{4.30b}
\end{equation}%
where $I_{0}$ is a constant. Since $B^{\ast }$ is of order unity, this shows
that $I(v^{\ast },c)\rightarrow I(\infty ,c)$ for $v^{\ast }\geq c>>1$ with
deviations of order $\exp (-v^{\ast 2})$. On this scale of velocities the
distribution function becomes
\begin{equation}
\phi (v^{\ast })\rightarrow \frac{I(\infty ,c)}{v^{3}}\exp \left( -\frac{2}{%
\zeta ^{\ast }}\int_{c}^{v^{\ast }}\frac{dx^{\prime }}{x^{\prime }}\nu
^{\ast }\left( x^{\prime }\right) \right) .  \label{4.30c}
\end{equation}%
This is not yet the exponential decay quoted in (\ref{2.12}) for the
Boltzmann equation. Instead, that exponential decay requires a still larger
velocity scale due to the slow approach of $\nu ^{\ast }\left( x^{\prime
}\right) $ in (\ref{4.30c}) to its large $x^{\prime }$ limiting form
\begin{equation}
\nu ^{\ast }\left( x^{\prime }\right) \rightarrow \beta _{G}x^{\prime }+%
\frac{\beta _{G}}{2x^{\prime }}+\text{ order }e^{-x^{\prime 2}},
\label{4.30d}
\end{equation}%
where $\beta _{G}/x(\alpha )=\beta _{B}$ is the coefficient of the large
velocity limit for the collision frequency given in (\ref{2.12c}). Thus (\ref%
{4.30c}) behaves as
\begin{equation}
\phi (v^{\ast })\rightarrow I(\infty,c)\exp(-(\frac{2}{\zeta^{\ast}}%
\beta_{G}(v^{\ast}-c)+3)\ln v^{\ast}+\frac{\beta_{G}}{\zeta^{\ast}}( \frac{1%
}{v^{\ast }}-\frac{1}{c})).  \label{4.31}
\end{equation}
The dominant exponential decay is the same as that for the Boltzmann
equation, (\ref{2.12}), with only the coefficient $\beta _{B}$ changed to $%
\beta _{G}$. In fact, a similar analysis of the derivation of that result
for the Boltzmann equation shows that the same intermediate velocity form (%
\ref{4.31}) applies there as well (with $\beta_{G}\rightarrow\beta_{B}$).
The cross over to pure exponential decay requires very large velocities. In
practical terms for $\alpha =0.8$ this form holds to within $0.1\%$ for $%
v^{\ast }\geq 6$ whereas the exponential decay has the same
accuracy only for much larger velocities. This is illustrated in
Figure 3 for $\alpha =0.8$. The derivative of the logarithm of the
distribution is plotted so that the initial slope for small
velocities  is near the Maxwellian value $(-2v)$, while the
asymptotic large velocity value is the constant coefficient of the
exponential decay $(\frac{-2\beta_{G}}{\zeta^{\ast}})$ shown as a
dotted line. The intermediate cross over is seen to be governed by
the asymptotic form (\ref{4.30c})
\begin{equation}
v^{\ast }\partial _{v^{\ast }}\ln \phi (v^{\ast })\sim -\frac{1}{v^{\ast }}%
\left( 3+\frac{2}{\zeta ^{\ast }}\nu ^{\ast }\left( v^{\ast
}\right) \right) .  \label{4.31a}
\end{equation}
Figure 3 also shows that this more general form persists to very
large velocities before the final exponential decay is attained.
This cross over form is expected to apply for the Boltzmann
equation as well and should be taken into account in simulation or
experimental attempts to measure the  over population at large
velocities.

 To explore the limiting form for $\alpha \rightarrow
1$ it is useful to integrate by parts in (\ref{4.27}) to get

\begin{eqnarray}
\phi (v^{\ast }) &=&\frac{A^{\ast }}{v^{\ast 3}}\int_{0}^{v^{\ast
}}dxx^{3}e^{-B^{\ast -1}x^{2}}\frac{d}{dx}\exp \left( -\frac{2}{\zeta ^{\ast
}}\int_{x}^{v^{\ast }}\frac{dx^{\prime }}{x^{\prime }}\nu ^{\ast }\left(
x^{\prime }\right) \right)  \nonumber \\
&=&A^{\ast }e^{-B^{\ast -1}v^{\ast 2}}-\frac{A^{\ast }}{v^{\ast 3}}%
\int_{0}^{v^{\ast }}dx\exp \left( -\frac{2}{\zeta ^{\ast }}\int_{x}^{v^{\ast
}}\frac{dx^{\prime }}{x^{\prime }}\nu ^{\ast }\left( x^{\prime }\right)
\right) \left( 3x^{2}-2B^{\ast -1}x^{4}\right) e^{-B^{\ast -1}x^{2}}.
\label{4.32}
\end{eqnarray}%
The second term on the right side vanishes at $\alpha =1$ leaving the
expected Maxwellian. For $\alpha <1$ the second term gives the exponential
decay at large velocities. In order to dominate the first term it is
necessary that $v^{\ast }>>1.$ Since the coefficient in the exponential
decay is proportional to $(1-\alpha )^{-1}$ the relevant domain for
overpopulation is $v^{\ast }>(1-\alpha )^{-1}$. Clearly for $\alpha
\rightarrow 1$ this overpopulation becomes physically insignificant.

In summary, it has been shown that all of the qualitative features of the
HCS for the hard sphere Boltzmann equation are reproduced by the Gaussian
model. In the next section it is shown that this quality of the model \
extends to transport properties as well.

\subsection{Limiting Case: Velocity independent collision frequency}

To emphasize the effects of the velocity dependence of the
collision frequency, it is instructive to consider the same
Gaussian model with a velocity independent collision frequency,
$\nu ^{\ast }\left( v^{\ast }\right) \rightarrow \nu _{\kappa
B}^{\ast }\left( \alpha \right) $. The HCS for the Gaussian model
then reduces to that of the BMD model. The single parameter of the
model, the constant collision frequency, can be chosen to fit the
shear viscosity or the thermal conductivity. Due to the choice of
$x(\alpha )$ made here the model is tailored to fit the thermal
conductivity. The functional form of $\nu _{\kappa B}^{\ast
}\left( \alpha \right) $ is given by (\ref{5.26b}) of the next
section. The HCS solution (\ref{4.81}) simplifies to

\begin{equation}
\phi (v^{\ast })=A^{\ast }\int_{0}^{\infty }dse^{-s}e^{-3s/p}e^{-B^{\ast
-1}e^{-2s/p}v^{\ast 2}},  \label{4.90}
\end{equation}
where $A^{\ast },B^{\ast },$ $M_{1}^{\ast }$, $M_{3}^{\ast }$ \ and $p$ are
given by
\begin{equation}
A^{\ast }=\left( \pi B^{\ast }\right) ^{-3/2},\hspace{0.25in}B^{\ast }=\frac{%
\left( p-2\right) }{p},\hspace{0.25in}M_{1}^{\ast }=M_{3}^{\ast }=\nu ^{\ast
},\hspace{0.25in}p=2\nu _{\kappa B}^{\ast }/\zeta ^{\ast }.  \label{4.100}
\end{equation}
A change of variables, $t=e^{-2s/p}v^{\ast 2}p/\left( p-2\right) $ allows
this to be expressed in terms of an incomplete gamma function
\begin{equation}
\phi (v^{\ast })=\frac{p}{2\pi ^{3/2}}\left( \frac{p-2}{p}\right)
^{p/2}v^{\ast -\left( 3+p\right) }\gamma \left( \frac{p+3}{2},\frac{p}{p-2}%
v^{\ast 2}\right),  \label{4.17}
\end{equation}
with
\begin{equation}
\gamma \left( x,y\right) =\int_{0}^{y}dte^{-t}t^{x-1}.  \label{4.18}
\end{equation}
Interestingly, the dimensionless distribution function is entirely
characterized by the single constant $p=2\nu _{\kappa B}^{\ast
}\left(
\alpha \right) /\zeta ^{\ast }\left( \alpha \right) $. Its relationship to $%
\alpha $ is fixed by the choices of cooling rate and collision frequency
\begin{equation}
p=\frac{8\left( 1+\frac{33}{16}\left( 1-\alpha \right) +\frac{19-3\alpha }{%
1024}c_{B}\left( \alpha \right) \right) }{5\left( 1-\alpha \right) \left( 1+%
\frac{3}{32}c_{B}(\alpha )\right) }.  \label{4.20}
\end{equation}

For small velocities the representation (\ref{2.7}) applies with $c\left(
\alpha \right) $ given by
\begin{equation}
c(\alpha )\text{ }\rightarrow \text{ }c_{G1}\left( \alpha \right) =\frac{8}{%
\left( p-4\right) p}.  \label{4.21}
\end{equation}%
Figure 1 shows a comparison of the coefficient $c_{G1}\left(
\alpha \right) $ with that for the Maxwell model given by
(\ref{2.16}). They are seen to be similar for weak dissipation but
the Gaussian model grows more rapidly with increasing dissipation.
Of course, this difference can be eliminated by a different choice
of the parameters for the Gaussian model for a closer agreement to
the Maxwell model rather than the hard sphere Boltzmann equation.
The accuracy of this polynomial representation is within a few
percent for relatively weak dissipation, comparable to that
observed for the hard sphere Boltzmann equation. It is clear from
this figure that the HCS for  models with velocity independent
collision frequencies differs from that of the Boltzmann equation
and the true Gaussian model at small velocities (see Figure 2).
Also, at small velocities, the HCS in (\ref{4.90}) can be
represented as a Gaussian given by
\begin{equation}
\phi (v^{\ast })\rightarrow \frac{p}{3+p}Ae^{-\frac{p(p+3)}{(p+5)(p-2)}%
v^{\ast 2}}. \label{4.21b}
\end{equation}%

The asymptotic behavior for large velocities is obtained from (\ref{4.17})
using the limiting form for the incomplete gamma function
\begin{equation}
\phi (v^{\ast })\rightarrow \frac{p}{2\pi ^{3/2}}\left( \frac{p}{p-2}\right)
^{-p/2}\Gamma \left( \frac{p+3}{2}\right) v^{\ast -\left( 3+p\right) }.
\label{4.22}
\end{equation}%
This algebraic decay is similar to that of the Maxwell model, and
in contrast to the exponential decay for the hard sphere Boltzmann
equation. This difference is due to neglect of the velocity
dependence of the collision frequency in both models. Since the
exact solution to the Gaussian model is known the crossover from
Gaussian to algebraic forms can be determined explicitly. Figure 4
illustrates this for $\alpha =0.8$. The cross over domain occurs
for $v^{\ast }\succsim 1$, increasing slightly with decreasing
$\alpha $, with no special intermediate behavior. Figure 5 shows a
comparison of the exponent for the algebraic decay for the
Gaussian model, $k(\alpha )=3+p\left( \alpha \right) $, with the
corresponding result for the Maxwell model obtained from the
solution to (\ref{2.18}). Near $\alpha =1 $ both exponents diverge
as $\left( 1-\alpha \right) ^{-1}$ but with different
coefficients.

\qquad In summary, the simple Gaussian model with constant
collision frequency captures semi-quantitatively all of the
relevant features of the Maxwell model. It has the additional
feature of demonstrating explicitly the solution for all
homogeneous states to show the rapid transition to the homogeneous
cooling state, and the detailed features of that state. The
algebraic decay at large velocities implies divergence of moments
of degree greater than some critical value for given $\alpha $.
The evolution of such moments for given initial states can be
studied in detail to characterize the growing over population of
large velocities in the HCS. However, the HCS for both the Maxwell
model and the Gaussian model with constant collision frequency
differ qualitatively from the Boltzmann result for large and small
velocities.

\section{Navier-Stokes Hydrodynamics}

In this section, states with smooth spatial and temporal
variations in the density, temperature and flow velocity are
considered. These are states for which a macroscopic hydrodynamic
description is expected to apply. First, the results of the
Chapman-Enskog method to solve the kinetic equation is recalled.
Next, the transport coefficients obtained from this solution are
identified exactly and in a first Sonine polynomial approximation.
Finally, these latter expressions are evaluated for the model and
compared with the corresponding results for the Boltzmann
equation.
\subsection{Chapman-Enskog solution}

The hydrodynamic equations for spatially inhomogeneous states are obtained
from a special solution to the kinetic equation generated by the
Chapman-Enskog method. The method is quite general and requires only the
properties (\ref{2.4}) for the collision operator. Since these are preserved
in the Maxwell and Gaussian models the results obtained earlier for the
Boltzmann equation \cite{BDKyS98} apply for the models as well. The solution
is ''normal'' in the sense that all space and time dependence occurs only
through the hydrodynamic fields. To first order in the spatial gradients of
these fields it is found to be
\begin{equation}
f\left( {\bf r},{\bf V},t\right) =f^{(0)}\left( {\bf r},{\bf V},t\right)
+f^{(1)}\left( {\bf r},{\bf V},t\right),  \label{5.1}
\end{equation}
where ${\bf V=v-u}\left( {\bf r},t\right) $ is the velocity relative to the
flow field. The first term of (\ref{5.1}) is the solution to the kinetic
equation to zeroth order in the spatial gradients
\begin{equation}
\frac{1}{2}\zeta ^{(0)}\left( {\bf r},t\right) \nabla _{{\bf V}}\cdot \left(
{\bf V}f^{(0)}\left( {\bf r},{\bf V},t\right) \right) =J[{\bf r},{\bf v}%
|f^{(0)}(t)].  \label{5.3}
\end{equation}
where the superscript on $\zeta ^{(0)}$ denotes (\ref{2.5}) evaluated with $%
f^{(0)}$. Equation (\ref{5.3}) is an equation for the velocity dependence of
$f^{(0)}\left( {\bf r},{\bf V},t\right) $ which is the {\em same} as that
for the HCS distribution of the previous sections. The dependence on ${\bf r}%
,t$ occurs only through the parameters of the HCS. More specifically, $%
f^{(0)}\left( {\bf r},{\bf V},t\right) $ is the {\em local} HCS obtained
from (\ref{2.6}) by replacing the density, temperature, and flow velocity by
their actual values in the spatially inhomogeneous state
\begin{equation}
f^{(0)}\left( {\bf r},{\bf V},t\right) =n\left( {\bf r},t\right) v_{0}^{-3}(%
{\bf r},t)\phi \left( V^{\ast }\right) ,\hspace{0.25in}V^{\ast }=V/v_{0}(%
{\bf r},t),\hspace{0.25in}v_{0}({\bf r},t)=\sqrt{2T({\bf r},t)/m}.
\label{5.2}
\end{equation}

The second term on the right side of (\ref{5.1}) is proportional to the
gradients
\begin{equation}
f^{(1)}\left( {\bf r},{\bf V},t\right) ={\cal A}\cdot \nabla \ln T\ +{\cal B}%
\cdot \nabla \ln n\ +{\cal C}_{ij}\frac{1}{2}\left( \partial
_{j}u_{i}+\partial _{i}u_{j}-\frac{2}{3}\delta _{ij}\nabla \cdot {\bf u}%
\right)  \label{5.5}
\end{equation}
(There are no contributions from the expansion of $\zeta \left( {\bf r}%
,t\right) $ to first order as this vanishes for both the Boltzmann case and
for the models). The functions ${\cal A}\left( {\bf V}|n,{\bf u},T\right) $,%
{\bf \ }${\cal B}\left( {\bf V}|n,{\bf u},T\right) $,{\bf \ } and ${\cal C}%
_{ij}\left( {\bf V}|n,{\bf u},T\right) $ are solutions to the integral
equations
\begin{equation}
\left( -\zeta ^{(0)}T\partial _{T}+{\cal L}-\frac{\zeta ^{(0)}}{2}\right)
{\cal A}={\bf A},  \label{5.6}
\end{equation}
\begin{equation}
\left( -\zeta ^{(0)}T\partial _{T}+{\cal L}\right) {\cal B}={\bf B}+\zeta
^{(0)}{\cal A},  \label{5.7}
\end{equation}
\begin{equation}
\left( -\zeta ^{(0)}T\partial _{T}+{\cal L}\right) {\cal C}_{ij}=C_{ij},
\label{5.8}
\end{equation}
with the definitions
\begin{equation}
{\bf A}({\bf V}|n,{\bf u},T)=\left( \frac{5}{2}+\frac{1}{2}{\bf V}%
\cdot \bbox {\nabla }_{V}\right)f^{(0)} {\bf V}-{\bf V}f^{(0)}-\frac{T}{m}%
\nabla _{V}f^{(0)},  \label{5.9}
\end{equation}
\begin{equation}
{\bf B}({\bf V}|n,{\bf u},T)=-{\bf V}f^{(0)}-\frac{T}{m}\bbox {\nabla}%
_{V}f^{(0)},  \label{5.10}
\end{equation}
\begin{equation}
C_{ij}({\bf V}|n,{\bf u},T)=V_{i}\left( \partial _{V_{j}}f^{(0)}\right) .
\label{5.11}
\end{equation}
The linear operator ${\cal L}$ is the collision operator expanded to first
order in $f^{(1)}$
\begin{equation}
{\cal L}f^{(1)}=\int d{\bf v}^{\prime }\frac{\delta J[{\bf r},{\bf v}|f(t)]}{%
\delta f\left( {\bf r},{\bf v}^{\prime },t\right) }\mid
_{f=f^{(0)}}f^{(1)}\left( {\bf r},{\bf V}^{\prime },t\right) .  \label{5.12}
\end{equation}

\subsection{Transport coefficients}

The Boltzmann equation and all models considered give the same macroscopic
balance equations for mass, energy, and momentum (or density, temperature,
and flow velocity) because they all imply the properties (\ref{2.4}). The
Navier-Stokes hydrodynamic equations follow by evaluating the momentum flux $%
P_{ij}$ and the heat flux $q$ in the macroscopic balance equations using the
Chapman-Enskog solution to first order in the spatial gradients, with the
results
\begin{equation}
P_{ij}=-\eta \left( \partial _{j}u_{i}+\partial _{i}u_{j}-\frac{2}{3}\delta
_{ij}\nabla \cdot {\bf u}\right) ,\hspace{0.3in}{\bf q}=-\kappa \nabla T-\mu
\nabla n.  \label{5.16}
\end{equation}
The first of these is Newton's viscosity law, where $\eta $ is the
shear viscosity. The second is a generalization of Fourier's law,
where $\kappa $ is the thermal conductivity. There is an
additional contribution for granular gases proportional to the
density gradient, with a transport coefficient $\mu $, that does
not occur for normal gases. These are identified from the
Chapman-Enskog solution as \cite{BDKyS98}

\begin{equation}
\eta =nT\left( \nu _{\eta }-\frac{1}{2}\zeta ^{(0)}\right) ^{-1},
\label{5.17}
\end{equation}
\begin{equation}
\kappa =\frac{5nT}{2m}(\nu _{\kappa }-2\zeta ^{(0)})^{-1}\left( 1+c\right),
\label{5.18}
\end{equation}
\begin{equation}
\mu =\frac{15T^{2}}{2m}\left( 2\nu _{\mu }-3\zeta ^{(0)}\right) ^{-1}\left(
\zeta ^{\ast (0)}\frac{\kappa }{\kappa _{0}}+\frac{1}{3}c\right).
\label{5.19}
\end{equation}
with the definitions
\begin{equation}
\nu _{\eta }=\frac{\int d{\bf V}\,D_{ij}({\bf V}){\cal LC}_{ij}({\bf V})}{%
\int d{\bf V}\,D_{ij}({\bf V}){\cal C}_{ij}({\bf V})},\hspace{0.3in}\nu
_{\kappa }=\frac{\int d{\bf V}\,{\bf S}({\bf V})\cdot {\cal LA}({\bf V})}{%
\int d{\bf V}\,{\bf S}({\bf V})\cdot {\cal A}({\bf V})},\quad \nu _{\mu }=%
\frac{\int d{\bf V}\,{\bf S}({\bf V})\cdot {\cal LB}({\bf V})}{\int d{\bf V}%
\,{\bf S}({\bf V})\cdot {\cal B}({\bf V})}.  \label{5.20}
\end{equation}
The functions $D_{ij}({\bf V})$ and ${\bf S}({\bf V})$ are
\begin{equation}
D_{ij}({\bf V})=m\left( V_{i}V_{j}-\frac{1}{3}V^{2}\delta _{ij}\right) ,%
\hspace{0.3in}{\bf S}({\bf V})={\bf V}\left( \frac{1}{2}mV^{2}-\frac{5}{2}%
T\right) .  \label{5.21}
\end{equation}
Also, $\kappa _{0}=15\eta _{0}/4m$ and $\eta _{0}=5\left( mT\right)
^{1/2}/16\sigma ^{2}\pi ^{1/2}$ are the low density values of the thermal
conductivity and the shear viscosity in the elastic limit, respectively. The
constant $c(\alpha )$ is the same as that occurring in the representation (%
\ref{2.7}), appropriate for either the Boltzmann equation or the model being
considered. The forms (\ref{5.17}) - (\ref{5.19}) provide the exact
expressions for these transport coefficients.

\subsection{Sonine Polynomial Approximation}

More explicit results require determination of $f^{(0)}$ and the solutions $%
{\cal A}$, ${\cal B}$, and ${\cal C}_{ij}$ to the linear integral equations (%
\ref{5.6})--(\ref{5.8}). The Gaussian model allows explicit construction of
these. However, in general it is useful to represent these quantities as an
expansion in a complete set of polynomials and generate approximations by
truncating the expansion. In practice the leading term in these expansions
provides a very accurate description over the full range of dissipation and
density. The determination of $f^{(0)}$ to leading order in the Sonine
polynomial has already been given by (\ref{2.7}). Similarly, the leading
contributions to the expansions of ${\cal A}$, ${\cal B}$, $\ $and ${\cal C}%
_{ij}$ are found to be \cite{BDKyS98}
\begin{equation}
\left(
\begin{array}{c}
{\cal A}(V) \\
{\cal B}(V) \\
{\cal C}_{ij}(V)%
\end{array}
\right) \rightarrow f_{M}(V)\left(
\begin{array}{c}
c_{{\cal A}}{\bf S}({\bf V}) \\
c_{{\cal B}}{\bf S}({\bf V}) \\
c_{{\cal C}}D_{ij}({\bf V})%
\end{array}
\right) ,\hspace{0.3in}f_{M}(V)=n\left( \pi v_{0}^{2}\right)
^{-3/2}e^{-\left( V/v_{0}\right) ^{2}}.  \label{5.22}
\end{equation}
with the coefficients
\begin{equation}
\left(
\begin{array}{c}
c_{{\cal A}} \\
c_{{\cal B}}%
\end{array}
\right) =\frac{2m}{15nT^{3}}\int d{\bf V}\left(
\begin{array}{c}
{\cal A}(V)\cdot {\bf S}({\bf V}) \\
{\cal B}(V)\cdot {\bf S}({\bf V})%
\end{array}
\right) =\left(
\begin{array}{c}
-\frac{2m}{5nT^{2}}\kappa \\
-\frac{2m}{5T^{3}}\mu%
\end{array}
\right) ,  \label{5.23}
\end{equation}
\begin{equation}
c_{{\cal C}}=\frac{T^{2}}{10n}\int d{\bf V}{\cal C}_{ij}(V)D_{ij}({\bf V})=-%
\frac{T^{2}}{n}\eta .  \label{5.24}
\end{equation}
The distribution function $f^{(1)}$ in this approximation is obtained from ( %
\ref{5.5})
\begin{equation}
f^{(1)}\rightarrow -f_{M}\left[ \frac{2m}{5nT^{3}}\left( \kappa \nabla T+\mu
\nabla n\right) \cdot {\bf S}({\bf V})+\frac{1}{nT^{2}}\eta \frac{1}{2}%
\left( \partial _{i}u_{j}+\partial _{j}u_{i}-\frac{2}{3}\delta _{ij}\nabla
\cdot {\bf u}\right) D_{ij}({\bf V})\right].  \label{5.25}
\end{equation}

To evaluate the transport coefficients the forms (\ref{5.17})--(\ref{5.19})
are used, with the frequencies $\nu _{\eta }(\alpha )$, $\nu _{\kappa
}(\alpha )$, and $\nu _{\mu }(\alpha )$ determined in this approximation by
\begin{equation}
\nu _{\eta }=\frac{\int d{\bf V}\,D_{ij}{\cal L}f_{M}D_{ij}}{\int d{\bf V}%
\,f_{M}D_{ij}D_{ij}},\quad \nu _{\kappa }=\nu _{\mu }=\frac{\int d{\bf V}\,%
{\bf S}\cdot {\cal L}f_{M}{\bf S}}{\int d{\bf V}f_{M}\,{\bf S}\cdot {\bf S}}.
\label{5.26}
\end{equation}%
These integrals have been calculated for the Boltzmann equation \cite%
{BDKyS98}
\begin{equation}
\nu _{\eta B}^{\ast }=\frac{\nu _{\eta B}}{\nu _{0}({\bf r},t)}=\left( 1-%
\frac{1}{4}\left( 1-\alpha \right) ^{2}\right) \left( 1-\frac{1}{64}%
c_{B}\left( \alpha \right) \right),  \label{5.26a}
\end{equation}%
\begin{equation}
\nu _{\kappa B}^{\ast }=\nu _{\mu B}^{\ast }=\frac{1+\alpha }{3}\left( 1+%
\frac{33}{16}\left( 1-\alpha \right) +\frac{19-3\alpha }{1024}c_{B}\left(
\alpha \right) \right).  \label{5.26b}
\end{equation}%
The average local frequency $\nu _{0}({\bf r},t)$ is given by (\ref{4.701}).
The corresponding results for the Maxwell model \cite{santos03} are
\begin{equation}
\nu _{\eta M}^{\ast }=\frac{\left( 1+\alpha \right) \left( 4-\alpha \right)
}{6}(1+\frac{3}{32}c_{B}(\alpha)), \label{5.26c}
\end{equation}%
\begin{equation}
\nu _{\kappa M}^{\ast }=\nu _{\mu M}^{\ast }=\frac{1}{24}\left( 1+\alpha
\right) \left( 19-11\alpha \right) (1+\frac{3}{32}c_{B}(\alpha)).
\label{5.26d}
\end{equation}%

Finally, it is straightforward to perform the same calculations
for the Gaussian model considered here. The form of the linearized
collision operator ${\cal L}$ for the Gaussian model is obtained
in Appendix D
\begin{equation}
{\cal L}f^{(1)}=\left( 1-{\cal P}\right) \nu f^{(1)}-\frac{5}{2}y\frac{\nu
g^{(0)}D_{ij}({\bf V})\int d{\bf v}^{\prime }D_{ij}f^{(1)}}{\int d{\bf v}%
D_{ij}D_{ij}g^{(0)}}.  \label{5.13}
\end{equation}%
Here ${\cal P}$ is a projection operator defined by
\begin{equation}
{\cal P}X=\nu g^{(0)}\psi _{\sigma }\int d{\bf v}\psi _{\sigma }X,\hspace{%
0.25in}g^{(0)}=g(\mid f^{(0)}),  \label{5.14}
\end{equation}%
and $\left\{ \psi _{\sigma }\right\} $ is the orthonormal set
\begin{equation}
\left(
\begin{array}{c}
\psi _{1} \\
{\bf \psi }_{2} \\
\psi _{3}%
\end{array}%
\right) =\left(
\begin{array}{c}
\sqrt{\frac{1}{N_{1}}}1 \\
\sqrt{\frac{3}{N_{2}}}{\bf V} \\
\sqrt{\frac{1}{N_{3}}}\left( V^{2}-\frac{N_{2}}{N_{1}}\right)%
\end{array}%
\right).  \label{5.15}
\end{equation}%
The normalization constants $N_{i}$ are given in Appendix D. \ With these
results the frequencies $\nu _{\eta G},$ $\nu _{\kappa G},$ and $\nu _{\mu
G} $ are found to be
\begin{equation}
\nu _{\eta G}=x\left( \frac{\int d{\bf V}\,D_{ij}\nu _{M}f_{M}D_{ij}}{\int d%
{\bf V}\,f_{M}D_{ij}D_{ij}}-\frac{1}{2}y\frac{\int d{\bf V}\,D_{ij}\nu
_{M}g^{(0)}D_{ij}}{\int d{\bf V}\,D_{ij}g^{(0)}D_{ij}}\right),  \label{5.15a}
\end{equation}%
\begin{equation}
\nu _{\kappa G}^{\ast }=\nu _{\mu G}^{\ast }=\frac{x\int d{\bf V}\,{\bf S}%
\cdot \left( 1-{\cal P}\right) \nu _{M}^{\ast }f_{M}{\bf S}}{\int d{\bf V}%
f_{M}\,S^{2}}.  \label{5.15b}
\end{equation}%
The constants $x\left( \alpha \right)$ and $y\left( \alpha \right) $ are now
chosen to assure accurate transport coefficients. This is most directly done
by requiring that the above frequencies are the same as those from the
Boltzmann equation, i.e.%
\begin{equation}
\nu _{\eta G}=\nu _{\eta B},\hspace{0.25in}\nu _{\kappa G}=\nu _{\kappa B}.
\label{5.15b2}
\end{equation}%
It follows form (\ref{5.17}) and (\ref{5.18}) that the Prandtl
number at $\alpha=1$ is $\nu_{\kappa}/\nu_{\eta}$. So this
choice assumes that the Gaussian model also will have the correct
Prandtl number in the elastic limit. This gives
\begin{equation}
x\left( \alpha \right) =\nu _{\kappa B}^{\ast }\left( \alpha \right) \frac{%
\int d{\bf V}f_{M}\,S^{2}}{\int d{\bf V}\,{\bf S}\cdot \left( 1-{\cal P}%
\right) \nu _{M}^{\ast }f_{M}{\bf S}},  \label{5.15d}
\end{equation}%
\begin{equation}
y\left( \alpha \right) =-2\left( \nu _{\eta B}\left( \alpha \right)
-x(\alpha )\frac{\int d{\bf V}\,D_{ij}\nu _{M}f_{M}D_{ij}}{\int d{\bf V}%
\,f_{M}D_{ij}D_{ij}}\right) \left( \frac{x(\alpha )\int d{\bf V}\,D_{ij}\nu
_{M}g^{(0)}D_{ij}}{\int d{\bf V}\,g^{(0)}D_{ij}D_{ij}}\right) ^{-1}.
\label{5.15c}
\end{equation}%
With these choices, the transport coefficients are given by (\ref{5.17})--(%
\ref{5.19}), and the only differences from the Boltzmann values results from
the replacement of $c_{B}\left( \alpha \right) $ by $c_{G}\left( \alpha
\right) $ in the expressions for $\kappa $ and $\mu $. It should be \ noted
that Eqs. (\ref{5.15c}) and (\ref{5.15d}) are implicit since the right sides
depend on $x\left( \alpha \right) $ through the collision frequency in Eq. (%
\ref{4.82}) that determines the parameters $A$ and $B_{ij}$. In practice the
calculation of $x(\alpha)$ is done iteratively. First, the integrals in (\ref%
{5.15d}) are evaluated at $\alpha =1$ to determine a zeroth order estimate
for $x\left( \alpha \right) $. Then (\ref{4.82}) is used to get a first
approximation to $A$ and $B_{ij}$. Next, these results are used in (\ref%
{5.15c}) and (\ref{5.15d}) to calculate the first approximation to $x\left(
\alpha \right) $ and $y\left( \alpha \right) $. The process is repeated
starting with this first approximation for $x\left( \alpha \right) .$ The
results reported here are for two iterations, showing \ good convergence of
the process. The fits obtained for $x$ and $y$ are
\[
x(\alpha)= 0.533 + 0.156\alpha -0.302\alpha^{2},
\]
\[
y(\alpha)=0.906 - 2.666\alpha + 0.724\alpha^{2}.
\]
For $\alpha =1$ these results reduce to
\begin{equation}
\nu _{\eta G}^{\ast }\left( 1\right) =\nu _{\eta B}^{\ast }\left( 1\right)
=1,\hspace{0.2in}\nu _{\kappa G}^{\ast }\left( 1\right) =\nu _{\mu G}^{\ast
}\left( 1\right) =\nu _{\kappa B}^{\ast }\left( 1\right) =\frac{2}{3},
\label{5.15g}
\end{equation}%
\begin{equation}
x\left( 1\right) =\allowbreak
\frac{448}{1153},\hspace{0.4cm}y\left( 1\right)
=-\frac{1247}{1106}. \label{5.15h}
\end{equation}%
These results define a new kinetic model for normal gases,
extending the ES model to one with a more realistic velocity
dependent collision frequency.

Also, for the special case of a constant collision frequency the general
results reduce to
\begin{equation}
x\left( \alpha \right) \nu _{M}^{\ast }=\nu _{\kappa B}^{\ast }\left( \alpha
\right) ,\hspace{0.4cm}y\left( \alpha \right) =-2\left( \frac{\nu _{\eta
B}^{\ast }\left( \alpha \right) }{\nu _{\kappa B}^{\ast }\left( \alpha
\right) }-1\right).  \label{5.15i}
\end{equation}%
This case is relevant also if one wanted to use the Gaussian kinetic model
to represent the Maxwell model. Then in (\ref{5.15i}) $\nu _{\eta B}^{\ast
}\left( \alpha \right) $ and $\nu _{\kappa B}^{\ast }\left( \alpha \right) $
should be replaced by $\nu _{\eta M}^{\ast }\left( \alpha \right) $ and $\nu
_{\kappa M}^{\ast }\left( \alpha \right) $, respectively. Finally, for both $%
\alpha =1$ and constant collision frequency the usual Ellipsoidal
Statistical model is recovered
\begin{equation}
\nu _{\eta G}^{\ast }\left( 1\right) =1,\hspace{0.2in}\nu _{\kappa G}^{\ast
}\left( 1\right) =\nu _{\mu G}^{\ast }\left( 1\right) =x\nu _{M}^{\ast }=%
\frac{2}{3},\hspace{0.4cm}y=-1.  \label{5.51k}
\end{equation}

Figures 6,7 and 8 show the shear viscosity, thermal conductivity
and the new transport coefficient $\mu$ for the various models
compared with the Boltzmann equation results. It is seen that the
shear viscosity for the Gaussian model with either a velocity
dependent or velocity independent collision frequency is
indistinguishable from the Boltzmann result. The small differences
between the velocity dependent collision frequency Gaussian model
and the Boltzmann results for the $\kappa$ and $\mu$ coefficients
are due to the differences between $c_{B}(\alpha )$ and
$c_{G}(\alpha )$, and by truncation of the above iteration
solution for $x\left( \alpha \right) $ after two steps. The
differences in the case of the constant collision frequency models
are more pronounced because as seen in Figure 1, the
$c_{G1}(\alpha)$ and $c_{M}(\alpha)$ are significantly different
from $c_{B}(\alpha)$ for smaller $\alpha$ values. These results
show that the Gaussian model has the ability to fit the transport
properties to the hard sphere Boltzmann results for all $\alpha$, 
including the correct Prandtl number $\eta C_{p}/\kappa =2/3$ at
$\alpha =1 $. The other model do not have this capacity and the
associated transport coefficients do not represent as well those
from the Boltzmann equation \cite{santos03}, although they yield
the correct Prandtl number at $\alpha =1$.  Clearly, the inclusion
of the velocity dependent collision frequency in the model allows
excellent agreement with the Boltzmann results.

\section{Hydrodynamic modes and Green-Kubo expressions}

The simplest solutions to the Navier-Stokes equations are those
for a large system with small perturbations about the HCS (not the
{\em local} HCS as considered above). The resulting five
independent solutions are referred to as hydrodynamic modes. For a
gas with elastic collisions, these would correspond to shear
diffusion, heat diffusion, and damped sound propagation. The
hydrodynamic modes are more complicated for inelastic collisions
but their properties have been worked out and
discussed \cite{BDKyS98}.

The Chapman-Enskog method provides a ''normal'' solution that implicitly
presumes the existence of a hydrodynamic description. A more fundamental
study of the context or validity of hydrodynamics is possible by determining
the possible solutions to the Boltzmann equation for small perturbations of
the HCS. The resulting linearized Boltzmann equation is obtained by
substituting $f=f_{hcs}\left[ 1+\Delta \right] $ into (\ref{2.1}) and
retaining terms linear in $\Delta $
\begin{equation}
\left( \partial _{s}+{\bf v}^{\ast }{\bf \cdot \nabla }^{\ast }+{\cal L}%
_{0}\right) \Delta =0.  \label{5.27}
\end{equation}
The dimensionless units of Section 4 have been used, and the linear operator
${\cal L}_{0}$ is defined by

\begin{equation}
{\cal L}_{0}\Delta =\phi ^{-1}{\cal L}\left( \phi \Delta \right) -\phi ^{-1}%
\frac{\zeta _{0}}{2}\frac{\partial }{\partial {\bf v}^{\ast }}\cdot \left(
\phi \Delta \right) .  \label{5.28}
\end{equation}
For elastic collisions the second term of (\ref{5.28}) vanishes, $\phi $
becomes the Maxwellian, and ${\cal L}_{0}$ is the usual linearized Boltzmann
collision operator. Its spectrum includes a five fold degenerate value at
zero. The corresponding eigenfunctions are $\Delta \rightarrow $ linear
combinations of $1,{\bf v}^{\ast },v^{\ast 2}$. These eigenfunctions are
known as the summational invariants because their sum for two particles is
conserved in a two particle collision. Thus, for $\alpha =1$ the
eigenfunctions and eigenvalues of ${\cal L}_{0}$ are known and constitute
the hydrodynamic modes in the long wavelength limit.

The identification of the linear combinations of these
hydrodynamic modes as eigenvalues and eigenfunctions of ${\cal
L}_{0}$ has been given recently \cite{DyB03,BDyR03} with the
results
\begin{equation}
{\cal L}_{0}\chi _{n}=\lambda _{n}\chi _{n},  \label{5.29}
\end{equation}
\begin{equation}
\lambda _{1}=0,\hspace{0.25in}\lambda _{2}=\frac{\zeta _{0}}{2},\hspace{%
0.25in}\lambda _{3}=\lambda _{4}=\lambda _{5}=-\frac{\zeta _{0}}{2}.
\label{5.30}
\end{equation}
The degeneracy for elastic collisions is partially broken, with some zero
eigenvalues going to $\pm \zeta _{0}/2$. The corresponding eigenfunctions
are
\begin{equation}
\chi _{1}=4+ v \partial _{v}\ln \phi \left( v^{\ast }\right) ,%
\hspace{0.25in}\chi _{2}=-3- v \partial _{v}\ln \phi \left(
v^{\ast }\right),  \label{5.31}
\end{equation}
\begin{equation}
{\bf \chi }_{n}=\widehat{v}_{n}\partial_{v}\ln \phi \left( v^{\ast }\right) ,\hspace{%
0.25in}n=3,4,5.  \label{5.32}
\end{equation}
For $\alpha =1$, $\partial _{v}\ln \phi \left( v^{\ast }\right) =-2{v%
}$ and the $\chi _{n}$ become linear combinations of $1,{\bf
v}^{\ast },v^{\ast 2}$. This suggests that
(\ref{5.30})-(\ref{5.32}) provide the hydrodynamic modes for
$\alpha <1$ as well. This is confirmed by noting that these
eigenvalues are the same as those of the macroscopic balance
equations in the long wavelength limit.

The velocity dependence of the hydrodynamic modes for $\alpha <1$
is complicated due to their definitions in terms of the HCS
distribution. An advantage of the Gaussian models is that this
distribution is known explicitly and the construction of the
eigenfunctions is straightforward. All of these modes are
characterized by $\partial_{v}\ln \phi$. This has already been
shown in Figure 3 for $\alpha=0.8$. Figure 9 shows the same data
but with the result for the velocity independent collision
frequency model included. The dashed curve in each case represents the elastic $%
\alpha =1$ limit. For the velocity independent collision frequency $\chi
_{1} $ approaches a constant for large $v$, due to the asymptotic algebraic
decay of $\phi $. For the velocity dependent collision frequency it
approaches $v$ according to the exponential decay of $\phi $.

Figure 9 shows that there are significant qualitative differences
from the hydrodynamic modes from the elastic limit when $v^{\ast}>
2$. This is the crossover of the distribution function to its
large velocity form (\ref{4.31a}). The fact that the hydrodynamic
modes are related to the log of the distribution function lends
new importance to these asymptotic forms.

Related properties are the fluxes appearing in the Green-Kubo
expressions for the transport coefficients. The expressions in the
previous section can be written in a form suggestive of Green-Kubo
relations for normal fluids \cite{DyB02}
\begin{equation}
\eta =\frac{nm\ell v_{0}(t)}{10}\int_{0}^{s}ds^{\prime }\left\langle
D_{ij}^{\ast }\Phi _{2,ij}^{\ast }(s^{\prime })\right\rangle e^{-\frac{1}{2}%
\zeta ^{\ast }s^{\prime }},  \label{5.33}
\end{equation}
\begin{equation}
\kappa =\frac{n\ell v_{0}(t)}{3}\int_{0}^{s}ds^{\prime }\left\langle {\bf S}%
^{\ast }\cdot {\bf \Phi }_{3}^{\ast }(s^{\prime })\right\rangle e^{\frac{1}{2%
}\zeta ^{\ast }s^{\prime }},  \label{5.34}
\end{equation}
\begin{equation}
\mu {\bf =}\frac{T}{n}\kappa +\frac{m\ell v_{0}^{3}(t)}{3}%
\int_{0}^{s}ds^{\prime }\left\langle {\bf S}^{\ast }\cdot \left( {\bf \Phi }%
_{1}^{\ast }(s^{\prime })-{\bf \Phi }_{3}^{\ast }(s^{\prime })\right)
\right\rangle . \label{5.35}
\end{equation}
The brackets denote an average over the HCS in the dimensionless velocities
\begin{equation}
\left\langle X\right\rangle =\int d{\bf v}^{\ast }\phi ({\bf v}^{\ast })X(%
{\bf v}^{\ast }).  \label{5.36}
\end{equation}
Furthermore, the dependence on the dimensionless time $s$ is defined by
\begin{equation}
X(s)=e^{s{\cal L}_{0}}X({\bf V}^{\ast }).  \label{5.37}
\end{equation}
The averages in these expressions therefore have the interpretation of time
correlation functions. The momentum flux $D_{ij}^{\ast }$ and heat flux $%
{\bf S}^{\ast }$ are the same as in (\ref{5.21}). They are fluxes in the
usual sense of the velocity $v$ \ times linear combinations of the
summational invariants $1,{\bf v}^{\ast },v^{\ast 2}$. In the elastic limit
the other functions ${\bf \Phi }_{n}^{\ast }$ also have these forms
\begin{equation}
{\bf \Phi }_{1}^{\ast }\rightarrow 0,\hspace{0.3in}\Phi _{2,ij}^{\ast
}\rightarrow D_{ij}^{\ast },\hspace{0.3in}{\bf \Phi }_{3}^{\ast }\rightarrow
{\bf S}^{\ast }({\bf V)}.  \label{5.38}
\end{equation}
The resulting expressions (\ref{5.33}) and (\ref{5.34}) for $\eta $ and $%
\kappa $ are then precisely the low density limits of the usual Green-Kubo
expressions as time integrals of flux autocorrelation functions \cite{Mc89}.

For $\alpha <1$ the functions ${\bf \Phi }_{i}^{\ast }$ are no longer simply
related to fluxes of the summational invariants. Instead they can be written
as fluxes for the hydrodynamic modes defined above
\begin{equation}
{\bf \Phi }_{1}^{\ast }={\bf v}^{\ast }\left( \chi _{1}+\chi _{2}\right) +
\frac{1}{2}{\bf {\chi}},  \label{5.39}
\end{equation}
\begin{equation}
{\bf \Phi }_{2,ij}^{\ast }=\frac{1}{2}\left( v_{i}\chi _{j}-\frac{1}{3}%
\delta _{ij}{\bf v\cdot} {\bf {\chi}}\right) ,\hspace{0.25in}j=3,4,5,
\label{5.40}
\end{equation}
\begin{equation}
{\bf \Phi }_{3}^{\ast }=\frac{1}{2}\left( {\bf v}^{\ast }\chi _{2}+{\bf {\chi%
} }\right),  \label{5.41}
\end{equation}
where ${\bf {\chi} }$ is the vector whose components are $\chi _{n}$, $%
n=3,4,5$. This relationship of the ''fluxes'' to the hydrodynamic
modes is the same as for a normal gas. Only the forms of the
hydrodynamic modes change for $\alpha <1$. However, since these
modes are significantly different at large velocities, it is
expected that their effect on the transport coefficients may be
important.

\section{Discussion}

The objective here has been to describe a simple but realistic
kinetic model for the hard sphere Boltzmann equation. The new
features of the Gaussian kinetic model defined in section 4
relative to previous models are 1) a velocity dependent collision
frequency, 2) two free parameters for a good description of
transport coefficients, and 3) applicability to both elastic and
inelastic collisions. For elastic collisions and constant
collision frequency it reduces to the ES kinetic model
\cite{ES66,Cergignani75}, while for inelastic collisions and
symmetric Gaussian it reduces to the BMD model \cite{BMyD96}. For
elastic collisions, constant collision frequency, and symmetric
Gaussian it becomes the usual BGK model \cite{Cergignani75}. It is
also shown here that the Gaussian model for constant collision
frequency can be ''tuned'' to represent well the more complicated
Maxwell models. One motivation for the generalization of a kinetic
model to include a velocity dependent collision frequency is a
more accurate description of the overpopulation at large
velocities for granular gases. The decrease of the distribution
function for large velocities in the simplest state of HCS is
algebraic for any model with a constant collision frequency,
including the Maxwell model. In contrast, the decay found from the
hard sphere Boltzmann equation is exponential due to the velocity
dependence of the loss term in the collision operator. This
qualitative difference may be important for driven states as well.
Although this asymptotic behavior occurs only for extremely large
velocities it can have an effect on the moments of the
distribution function. In addition it has been shown in Section 6
that the hydrodynamic modes and the Green-Kubo fluxes depend on
the log of the HCS distribution function, so this asymptotic
behavior is even more important. The Gaussian model with velocity
dependent collision frequency incorporates this behavior and in
addition gives a quite good quantitative representation of the HCS
distribution function for small velocities as well. This is
illustrated in Fig 1 where $c_{G}(\alpha)$ shows significant
improvement over the velocity independent case. As a consequence
the transport coefficients $\kappa (\alpha)$ and $\mu (\alpha)$
are also significantly improved due to their dependence on
$c_{G}(\alpha)$.

The second feature of a non-symmetric Gaussian provides an additional
parameter beyond the collision frequency that can be chosen to optimize the
quality of all transport coefficients. Here they are chosen such that the
shear viscosity is accurate for both the constant collision frequency and
the velocity dependent collision frequency for all values of the restitution
coefficient. The other transport coefficients are accurate in the elastic
limit, including the correct Prandtl number for both cases. For inelastic
collisions the agreement with Boltzmann remains excellent for the velocity
dependent collision frequency case for all $\alpha$. This is a primary
improvement of the Gaussian model. In contrast, the transport coefficients
from the Maxwell model are quite different from those of the hard sphere
Boltzmann equation, and the other kinetic models using a symmetric Gaussian
all give the wrong Prandtl number.

An advantage of most kinetic models is their structural simplicity. They can
be solved exactly for many states as functionals of a few moments of the
distribution. These moments still obey complicated nonlinear integral
equations but the problem is simplified to the extent that exact results are
often possible for states with sufficient symmetry. An example is given here
for homogeneous states where the exact solution is obtained in terms of the
parameters of the Gaussian gain term, $A\left( s\right) $ and $B_{ij}\left(
s\right) ,$ which in turn are defined in terms of the moments $M_{\lambda
}\left( s\right) $. It is shown that an arbitrary homogeneous initial
condition evolves after a few collisions to a universal scaling solution,
the HCS. Such behavior is expected also from the hard sphere Boltzmann
equation but its complexity has precluded a proof to date. It is useful also
to have the explicit representation of the HCS for other purposes as well.
Here it has been noted that the hydrodynamic modes for weakly inhomogeneous
states are described by eigenfunctions of the linearized Boltzmann collision
operator. These eigenfunctions are determined from the HCS and a contrast
with the corresponding eigenfunctions for elastic collisions was made
possible by the explicit results for the HCS for the kinetic model.
Significant differences are observed between the cases of the velocity
independent and velocity dependent collision frequency, due to the
qualitative differences in the large velocity dependences of the HCS. This
is also related to the Hilbert space for formulating the eigenvalue problem
for the linearized kinetic equation. The natural scalar product is an
integration over the velocities weighted by the HCS distribution function.
Due to the algebraic decay at large velocities for the constant collision
frequency case (including the Maxwell model), polynomials of high degree do
not exist in this space \cite{DyB03}. This restriction does not occur for
the hard sphere Boltzmann equation or the Gaussian model with velocity
dependent collision frequency. It is of some interest to study any
qualitative differences in the spectrum of the linearized collision operator
and any consequences for the existence of hydrodynamics. The Gaussian
kinetic model provides a tractable context to address this issue.

The most interesting states for experimental purposes are quasi-steady
states for systems driven at the boundaries. For states of high spatial
symmetry the kinetic model again offers the advantage of an exact solution
as a functional of low degree moments. An example is that of uniform shear
flow where an exact solution for the distribution function has been obtained in the case of a 
symmetric Gaussian \cite{BRyM97}. The result
applies even for large shear rates so the rheology of states far from
equilibrium can be studied directly. The Gaussian model described here also
can be solved exactly for uniform shear flow and will be given elsewhere.
Vibrated systems, with and without gravity, have been studied on the basis
of the Boltzmann equation using Monte Carlo simulation methods leading to a
number of important results bearing on experiments (e.g., boundary layers %
\cite{BRMyM00}, dependence of velocity distribution on the distance from the
driving wall \cite{ByRM}, symmetry breaking \cite{BRMMyG2002}). The Gaussian
kinetic model may be simple enough for a complementary analytical study of
such problems.

In summary, the work here has extended earlier kinetic models to bring
closer correspondence with the Boltzmann equation for the HCS and small
spatial perturbations of that state. The price for these improvements is an
increased complexity of the model, although this has not been an impediment
for the simple states considered here. It remains to demonstrate significant
new results for more complex states, not already addressed by the simpler
existing models.

\section{Acknowledgments}

This research was supported in part by Department of Energy grant
DE-FG02-02ER54677 and by grants from Asociaci\'{o}n Mexicana de
Cultura, A.C and from Consejo Nacional de Ciencia y Tecnolog\'{i}a
grant 020163. The authors are grateful to A. Santos for sending
results prior to publication and for his careful reading of the
first draft. J.W.D and A.B acknowledge the hospitality of the
University of Seville where part of this work was performed.

\appendix

\section{Motivation for Gaussian model}

\label{appA}The Gaussian model results from an approximation to the gain
contribution to the Boltzmann collision operator, denoted by $g({\bf r},{\bf %
V},t\mid f)$ in (\ref{3.7}). The specific choice of a Gaussian can be
interpreted as resulting from maximizing the information entropy $I[g]$
\begin{equation}
I[g]=\int d{\bf \mu }\left( {\bf v}\right) g({\bf r},{\bf V},t\mid f)\ln g(%
{\bf r},{\bf V},t\mid f),\hspace{0.25in}d{\bf \mu }\left( {\bf v}\right) =d%
{\bf v}\nu ({\bf v}),  \label{a.1}
\end{equation}
among the class of functions whose weighted moments of degree $2$ are
specified
\begin{equation}
\int d{\bf \mu }\left( {\bf v}\right) \left(
\begin{array}{c}
1 \\
{\bf v} \\
\frac{1}{2}mV^{2}%
\end{array}
\right) g({\bf r},{\bf V},t\mid f)=\left(
\begin{array}{c}
G_{1} \\
{\bf G}_{2} \\
G_{3ij}%
\end{array}
\right) .  \label{a.2}
\end{equation}
The measure for the velocity integration has been chosen to include the
velocity dependence of the collision frequency. Incorporating these
constraints with \ Lagrange multiplyers and minimizing $I[g]$ leads directly
to the Gaussian form
\begin{equation}
g({\bf r},{\bf V},t\mid f)=\exp \left( -\lambda_{1}-{\bf \lambda}_{2}\cdot
{\bf v-}\lambda_{3ij}v_{i}v_{j}\right),  \label{a.3}
\end{equation}
where the coefficients $\lambda_\alpha$ are determined in terms of $G_\alpha$
from (\ref{a.2}). Thus, if the only known or important exact properties of
the gain term are the moments in (\ref{a.2}) then (\ref{a.3}) is a
''natural'' choice for the model.

It may be useful to recall that $g({\bf r}, {\bf V},t|f)$ is
exactly Gaussian for $f=$ Maxwellian at $\alpha = 1$. It has been
verified numerically that this property remains true to an
excellent approximation for $\alpha<1$ as well, with only the
parameters of the Gaussian changing. This gives further support
for the choice (\ref{a.3}).

\section{Formal solution for homogeneous states}

\label{appB}The formal solution to the Gaussian model kinetic equation (\ref%
{4.75}) is
\begin{eqnarray}
f^{\ast }(v^{\ast },s) &=&e^{-\left( \frac{1}{2}\zeta ^{\ast }\left( 3+{\bf v%
}^{\ast }\cdot \nabla _{{\bf v}^{\ast }}\right) +\nu ^{\ast }(v^{\ast
})\right) s}f^{\ast }(v^{\ast },0)  \nonumber \\
&&+\int_{0}^{s}ds^{\prime }e^{-\left( \frac{1}{2}\zeta ^{\ast }\left( 3+{\bf %
v}^{\ast }\cdot \nabla _{{\bf v}^{\ast }}\right) +\nu ^{\ast }(v^{\ast
})\right) \left( s-s^{\prime }\right) }\nu ^{\ast }\left( v^{\ast }\right)
g^{\ast }\left( v^{\ast },s^{\prime }\right).  \label{a1}
\end{eqnarray}
The action of the exponential in (\ref{a1}) can be determined as follows.
Define a function $X(v^{\ast },s)$ by
\begin{equation}
X(v^{\ast },s)=e^{-\left( \frac{1}{2}\zeta ^{\ast }\left( 3+{\bf v}^{\ast
}\cdot \nabla _{{\bf v}^{\ast }}\right) +\nu ^{\ast }(v^{\ast })\right)
s}X\left( v^{\ast }\right),  \label{a2}
\end{equation}
which then obeys the equation
\begin{equation}
\left( \partial _{s}+\frac{1}{2}\zeta ^{\ast }\left( 3+{\bf v}^{\ast }\cdot
\nabla _{{\bf v}^{\ast }}\right) +\nu ^{\ast }(v^{\ast })\right) X=0.
\label{a3}
\end{equation}
Next introduce
\begin{equation}
X(v^{\ast },s)=e^{-\frac{1}{2}\zeta ^{\ast }s{\bf v}^{\ast }\cdot \nabla _{%
{\bf v}^{\ast }}}\overline{X}(v^{\ast },s),  \label{a4}
\end{equation}
so that $\overline{X}(v^{\ast },s)$ obeys the equation\
\begin{equation}
\left( \partial _{s}+\frac{3}{2}\zeta ^{\ast }+e^{\frac{1}{2}\zeta ^{\ast }s%
{\bf v}^{\ast }\cdot \nabla _{{\bf v}^{\ast }}}\nu ^{\ast }(v^{\ast })e^{-%
\frac{1}{2}\zeta ^{\ast }s{\bf v}^{\ast }\cdot \nabla _{{\bf v}^{\ast
}}}\right) \overline{X}=0.  \label{a5}
\end{equation}
From the identity
\begin{equation}
e^{\frac{1}{2}\zeta ^{\ast }s{\bf v}^{\ast }\cdot \nabla _{{\bf v}^{\ast
}}}F(v^{\ast })e^{-\frac{1}{2}\zeta ^{\ast }s{\bf v}^{\ast }\cdot \nabla _{%
{\bf v}^{\ast }}}=F\left( e^{\frac{1}{2}\zeta ^{\ast }s}v^{\ast }\right),
\label{a6}
\end{equation}
this equation becomes
\begin{equation}
\left( \partial _{s}+\frac{3}{2}\zeta ^{\ast }+\nu ^{\ast }\left( e^{\frac{1%
}{2}\zeta ^{\ast }s}v^{\ast }\right) \right) \overline{X}=0.  \label{a7}
\end{equation}
This can be integrated directly and inserted \ in (\ref{a4}) to give
\begin{eqnarray}
X\left( v^{\ast },s\right) &=&e^{-\frac{3}{2}\zeta ^{\ast }s}e^{-\frac{1}{2}%
\zeta ^{\ast }s{\bf v}^{\ast }\cdot \nabla _{{\bf v}^{\ast }}}\exp \left(
-\int_{0}^{s}ds^{\prime }\nu ^{\ast }\left( e^{\frac{1}{2}\zeta ^{\ast
}s^{\prime }}v^{\ast }\right) \right) X\left( v^{\ast }\right)  \nonumber \\
&=&e^{-\frac{3}{2}\zeta ^{\ast }s}\exp \left( -\int_{0}^{s}ds^{\prime }\nu
^{\ast }\left( e^{-\frac{1}{2}\zeta ^{\ast }s^{\prime }}v^{\ast }\right)
\right) X\left( e^{-\frac{1}{2}\zeta ^{\ast }s}v^{\ast }\right).  \label{a8}
\end{eqnarray}
The formal solution to the kinetic equation becomes
\begin{eqnarray}
f^{\ast }(v^{\ast },s) &=&e^{-\frac{3}{2}\zeta ^{\ast }s}K(v^{\ast
},s)f^{\ast }(e^{-\frac{1}{2}\zeta ^{\ast }s}v^{\ast },0)  \nonumber \\
&&+\int_{0}^{s}ds^{\prime }e^{-\frac{3}{2}\zeta ^{\ast }s^{\prime
}}K(v^{\ast },s^{\prime })\nu ^{\ast }\left( e^{-\frac{\zeta ^{\ast }}{2}%
s^{\prime }}v^{\ast }\right) g^{\ast }\left( e^{\frac{-\zeta ^{\ast }}{2}%
s^{\prime }}v^{\ast },s-s^{\prime }\right),  \label{a9}
\end{eqnarray}
\begin{equation}
K(v^{\ast },s)=\exp \left( -\int_{0}^{s}ds^{\prime }\nu ^{\ast }\left( e^{-%
\frac{1}{2}\zeta ^{\ast }s^{\prime }}v^{\ast }\right) \right).  \label{a10}
\end{equation}
It is interesting to note that no use of the explicit form for
$g^{\ast}$ has been used. So, this result applies to the Boltzmann
equation as well.

\section{Moment conditions}

\label{appC}The HCS for the Gaussian model is given by (\ref{4.81})
\begin{equation}
\phi (v^{\ast })=\int_{0}^{\infty }ds^{\prime }e^{-\frac{3}{2}\zeta ^{\ast
}s^{\prime }}K(v^{\ast },s^{\prime })\nu ^{\ast }(e^{-\frac{1}{2}\zeta
^{\ast }s^{\prime }}v^{\ast })g^{\ast }(e^{-\frac{1}{2}\zeta ^{\ast
}s^{\prime }}v^{\ast },\infty ).  \label{b.1}
\end{equation}
This is restricted by the moment conditions (\ref{4.77})
\begin{equation}
\left(
\begin{array}{c}
M_{1}^{\ast } \\
M_{3}^{\ast }%
\end{array}
\right) =\int d{\bf v}^{\ast }\left(
\begin{array}{c}
1 \\
\frac{2}{3}v^{\ast 2}%
\end{array}
\right) \nu ^{\ast }(v^{\ast })\phi ({\bf v}^{\ast }).  \label{b2}
\end{equation}
These conditions can be verified by direct integration
\[
\left(
\begin{array}{c}
M_{1}^{\ast } \\
M_{3}^{\ast }%
\end{array}
\right) =\int d{\bf v}^{\ast }\left(
\begin{array}{c}
1 \\
\frac{2}{3}v^{\ast 2}%
\end{array}
\right) \nu ^{\ast }(v^{\ast })\int_{0}^{\infty }ds^{\prime }e^{-\frac{3}{2}%
\zeta ^{\ast }s^{\prime }}K(v^{\ast },s^{\prime })\nu ^{\ast }(e^{-\frac{1}{2%
}\zeta ^{\ast }s^{\prime }}v^{\ast })g^{\ast }(e^{-\frac{1}{2}\zeta ^{\ast
}s^{\prime }}v^{\ast },\infty )
\]
\[
=\int d{\bf v}^{\ast }\left(
\begin{array}{c}
1 \\
\frac{2}{3}v^{\ast 2}%
\end{array}
\right) \frac{1}{v^{\ast 3}}\int_{0}^{v^{\ast }}dxx^{2}\nu ^{\ast
}(x)g^{\ast }(x,\infty )\frac{2}{\zeta ^{\ast }}\nu ^{\ast }(v^{\ast })%
\overline{K}(v^{\ast },x),
\]
with the notation
\[
\overline{K}(v^{\ast },x)\equiv \exp \left( -\frac{2}{\zeta ^{\ast }}%
\int_{x}^{v^{\ast }}\frac{dx^{\prime }}{x^{\prime }}\nu ^{\ast }\left(
x^{\prime }\right) \right).
\]
Next eliminate $2\nu ^{\ast }(v^{\ast })/\zeta ^{\ast }$ by noting it can be
generated by differentiating $\overline{K}$
\[
\left(
\begin{array}{c}
M_{1}^{\ast } \\
M_{3}^{\ast }%
\end{array}
\right) =-4\pi \int_{0}^{\infty }dv^{\ast }\left(
\begin{array}{c}
1 \\
\frac{2}{3}v^{\ast 2}%
\end{array}
\right) \int_{0}^{v^{\ast }}dxx^{2}\nu ^{\ast }(x)g^{\ast }(x,\infty )\frac{d%
\overline{K}(v^{\ast },x)}{dv^{\ast }}
\]
\[
=-4\pi \int_{0}^{\infty }dxx^{2}\nu ^{\ast }(x)g^{\ast }(x,\infty
)\int_{x}^{\infty }dv^{\ast }\left(
\begin{array}{c}
1 \\
\frac{2}{3}v^{\ast 2}%
\end{array}
\right) \frac{d\overline{K}(v^{\ast },x)}{dv^{\ast}}
\]
\[
=4\pi \int_{0}^{\infty }dxx^{2}\nu ^{\ast }(x)g^{\ast }(x,\infty )\left\{
\left(
\begin{array}{c}
1 \\
\frac{2}{3}x^{2}%
\end{array}
\right) +\int_{x}^{\infty }dv^{\ast }\left(
\begin{array}{c}
0 \\
\frac{4}{3}v^{\ast }%
\end{array}
\right) \overline{K}(v^{\ast },x)\right\}
\]
\begin{equation}
=\left(
\begin{array}{c}
M_{1}^{\ast } \\
M_{3}^{\ast }-\zeta ^{\ast }%
\end{array}
\right) +4\pi \int_{0}^{\infty }dv^{\ast }\left(
\begin{array}{c}
0 \\
\frac{4}{3}v^{\ast }%
\end{array}
\right) \int_{0}^{v^{\ast }}dxx^{2}\nu ^{\ast }(x)g^{\ast }(x,\infty )%
\overline{K}(v^{\ast },x) . \label{b.3}
\end{equation}
where use has been \ made of (\ref{4.77a}). The second term of (\ref{b.3})
can be recognized as moments of $\phi (v^{\ast })$ to give the desired
result
\begin{eqnarray}
\left(
\begin{array}{c}
M_{1}^{\ast } \\
M_{3}^{\ast }%
\end{array}
\right) &=&\left(
\begin{array}{c}
M_{1}^{\ast } \\
M_{3}^{\ast }-\zeta ^{\ast }%
\end{array}
\right) +\frac{\zeta ^{\ast }}{2}\int d{\bf v}^{\ast }\left(
\begin{array}{c}
0 \\
\frac{4}{3}v^{\ast 2}%
\end{array}
\right) \phi (v^{\ast })  \nonumber \\
&=&\left(
\begin{array}{c}
M_{1}^{\ast } \\
M_{3}^{\ast }%
\end{array}
\right) .  \label{b3a}
\end{eqnarray}
The last equality follows from (\ref{2.4a}) in the form
\begin{equation}
\left(
\begin{array}{c}
1 \\
1%
\end{array}
\right) =\int d{\bf v}^{\ast }\left(
\begin{array}{c}
1 \\
\frac{2}{3}v^{\ast 2}%
\end{array}
\right) \phi ({\bf v}^{\ast }).  \label{b4}
\end{equation}
This confirms the consistency of moment conditions (\ref{4.77}) and (\ref%
{4.77a}).

The two equations (\ref{b4}) fix the values of $A^{\ast }$ and $B^{\ast-1 }$
in the Gaussian model (\ref{4.75}). A convenient representation is
\[
\frac{3}{2}=\frac{\int_{0}^{\infty }dv^{\ast }v^{\ast }\int_{0}^{v^{\ast
}}dxx^{2}\nu ^{\ast }(x)e^{-B^{\ast-1 }x^{2}}\overline{K}(v^{\ast },x)}{%
\int_{0}^{\infty }dv^{\ast }v^{\ast -1}\int_{0}^{v^{\ast }}dxx^{2}\nu ^{\ast
}(x)e^{-B^{\ast-1 }x^{2}}\overline{K}(v^{\ast },x)},
\]%
or
\begin{equation}
0=\int_{0}^{\infty }dv^{\ast }\left( v^{\ast }-\frac{3}{2}v^{\ast -1}\right)
\int_{0}^{v^{\ast }}dxx^{2}\nu ^{\ast }(x)e^{-B^{\ast-1 }x^{2}}\overline{K}%
(v^{\ast },x).  \label{b.5}
\end{equation}%
This determines $B^{\ast }$. Next $A^{\ast }$ is obtained from
\begin{equation}
A^{\ast -1}=\frac{8\pi }{\zeta ^{\ast }}\int_{0}^{\infty }dv^{\ast }v^{\ast
-1}\int_{0}^{v^{\ast }}dxx^{2}\nu ^{\ast }(x)e^{-B^{\ast-1 }x^{2}}\overline{K%
}(v^{\ast },x).  \label{b.6}
\end{equation}%
Finally, with $A^{\ast }$ and $B^{\ast-1 }$ known, the moments are
determined from (\ref{4.77a})
\begin{equation}
\left(
\begin{array}{c}
M_{1}^{\ast } \\
M_{3}^{\ast }-\zeta ^{\ast }%
\end{array}%
\right) =\int d{\bf v}^{\ast }\left(
\begin{array}{c}
1 \\
\frac{2}{3}v^{\ast 2}%
\end{array}%
\right) \nu ^{\ast }(v^{\ast })A^{\ast }e^{-B^{\ast-1 }v^{\ast 2}}.
\label{b.7}
\end{equation}

\section{Linearized collision operator}

\label{appD}The collision operator for the Gaussian model is
\begin{equation}
J\left( [{\bf r},{\bf v}|f(t)\right) \equiv -\nu ({\bf r},v,t)\left[ f\left(
{\bf r},{\bf V},t\right) -g({\bf r},{\bf V},t\mid f)\right].  \label{c.1}
\end{equation}
The distribution function is expanded as
\begin{equation}
f\left( {\bf r},{\bf V},t\right) =f^{(0)}\left( {\bf r},{\bf V},t\right)
+f^{(1)}\left( {\bf r},{\bf V},t\right) +.. , \label{c.2}
\end{equation}
\begin{equation}
g({\bf r},{\bf V},t\mid f)=g({\bf r},{\bf V},t\mid f^{(0)})+\int d{\bf v}%
^{\prime }\frac{\delta g({\bf r},{\bf V},t\mid f)}{\delta f\left( {\bf r},%
{\bf v}^{\prime },t\right) }\mid _{f=f^{(0)}}f^{(1)}\left( {\bf r},{\bf V}%
^{\prime },t\right) +.. . \label{c.3}
\end{equation}
The linearized collision operator is therefore
\begin{eqnarray}
{\cal L}f^{(1)} &=&-\int d{\bf v}^{\prime }\frac{\delta J[{\bf r},{\bf v}%
|f(t)]}{\delta f\left( {\bf r},{\bf v}^{\prime },t\right) }\mid
_{f=f^{(0)}}f^{(1)}\left( {\bf r},{\bf V}^{\prime },t\right)  \nonumber \\
&=&\nu ({\bf r},v,t)\left[ f^{(1)}\left( {\bf r},{\bf V},t\right) -\int d%
{\bf v}^{\prime }\frac{\delta g({\bf r},{\bf V},t\mid f)}{\delta f\left(
{\bf r},{\bf v}^{\prime },t\right) }\mid _{f=f^{(0)}}f^{(1)}\left( {\bf r},%
{\bf V}^{\prime },t\right) \right].  \label{c.4}
\end{eqnarray}
The second term of (\ref{c.3}) can be made more explicit by recalling that
the functional dependence of $g({\bf r},{\bf V},t\mid f)$ occurs only
through $A$ and $B_{ij}$%
\[
\int d{\bf v}^{\prime }\frac{\delta g({\bf r},{\bf V},t\mid f)}{\delta
f\left( {\bf r},{\bf v}^{\prime },t\right) }\mid _{f=f^{(0)}}f^{(1)}\left(
{\bf r},{\bf V}^{\prime },t\right) =g({\bf r},{\bf V},t\mid f^{(0)})\left[
\int d{\bf v}^{\prime }\frac{\delta \ln A({\bf r},t\mid f)}{\delta f\left(
{\bf r},{\bf v}^{\prime },t\right) }\mid _{f=f^{(0)}}f^{(1)}\left( {\bf r},%
{\bf V}^{\prime },t\right) \right.
\]
\begin{equation}
\left. -V_{i}V_{j}\int d{\bf v}^{\prime }\frac{\delta B_{ij}^{-1}({\bf r}%
,t\mid f)}{\delta f\left( {\bf r},{\bf v}^{\prime },t\right) }\mid
_{f=f^{(0)}}f^{(1)}\left( {\bf r},{\bf V}^{\prime },t\right) \right].
\label{c.5}
\end{equation}
Using the definition of $B_{ij}$ in (\ref{4.8}) gives
\begin{eqnarray}
\frac{\delta B_{ij}^{-1}({\bf r},t\mid f)}{\delta f\left( {\bf r},{\bf v}%
^{\prime },t\right) } &=&\left( -B^{-1}({\bf r},t\mid f^{(0)})\frac{\delta B(%
{\bf r},t\mid f)}{\delta f\left( {\bf r},{\bf v}^{\prime },t\right) }B^{-1}(%
{\bf r},t\mid f^{(0)})\right) _{ij}  \nonumber \\
&=&-\frac{9}{B_{kk}^{2}}\left( \frac{1}{3}\frac{\delta B_{kk}({\bf r},t\mid
f)}{\delta f\left( {\bf r},{\bf v}^{\prime },t\right) }\delta _{ij}+\frac{%
y(\alpha)}{nm}D_{ij}({\bf V})\right) .  \label{c.5a}
\end{eqnarray}
Then (\ref{c.5}) becomes
\[
\int d{\bf v}^{\prime }\frac{\delta g({\bf r},{\bf V},t\mid f)}{\delta
f\left( {\bf r},{\bf v}^{\prime },t\right) }\mid _{f=f^{(0)}}f^{(1)}\left(
{\bf r},{\bf V}^{\prime },t\right) =g({\bf r},{\bf V},t\mid f^{(0)})\left[
\int d{\bf v}^{\prime }\frac{\delta \ln A({\bf r},t\mid f)}{\delta f\left(
{\bf r},{\bf v}^{\prime },t\right) }\mid _{f=f^{(0)}}f^{(1)}\left( {\bf r},%
{\bf V}^{\prime },t\right) \right.
\]
\begin{equation}
\left. +V^{2}\frac{3}{B_{kk}^{2}}\int d{\bf v}^{\prime }\frac{\delta B_{kk}(%
{\bf r},t\mid f)}{\delta f\left( {\bf r},{\bf v}^{\prime },t\right) }\mid
_{f=f^{(0)}}f^{(1)}\left( {\bf r},{\bf V}^{\prime },t\right) +\frac{%
9y(\alpha)}{nm^{2}B_{kk}^{2}}D_{ij}({\bf V})\int d{\bf v}^{\prime }D_{ij}(%
{\bf V}^{\prime })f^{(1)}\left( {\bf r},{\bf V}^{\prime },t\right) \right].
\label{c.5b}
\end{equation}
The expansion (\ref{c.2}) leads to an corresponding expansion for the
moments
\begin{eqnarray}
\left(
\begin{array}{c}
M_{1} \\
{\bf M}_{2} \\
M_{3}%
\end{array}
\right) &=&\int d{\bf v}\left(
\begin{array}{c}
1 \\
{\bf v} \\
\frac{1}{2}m\left( {\bf v}-{\bf u}\right) ^{2}%
\end{array}
\right) \nu ({\bf r},v,t)\left( f^{(0)}\left( {\bf r},{\bf V},t\right)
+f^{(1)}\left( {\bf r},{\bf V},t\right) +..\right)  \nonumber \\
&=&\left(
\begin{array}{c}
M_{1}^{(0)} \\
{\bf M}_{2}^{(0)} \\
M_{3}^{(0)}%
\end{array}
\right) +\left(
\begin{array}{c}
M_{1}^{(1)} \\
{\bf M}_{2}^{(1)} \\
M_{3}^{(1)}%
\end{array}
\right) +...  \label{c.6}
\end{eqnarray}
The coefficients $A({\bf r},t\mid f^{(0)})$ and $B_{ij}({\bf r},t\mid
f^{0})=\left( B_{kk}({\bf r},t\mid f^{0})/3\right) \delta _{ij}=B\delta
_{ij} $ are determined from $M_{\lambda }^{(0)}$ just as is done in Appendix
B. The remaining part of the moment conditions are
\begin{equation}
\left(
\begin{array}{c}
M_{1}^{(1)} \\
{\bf M}_{2}^{(1)} \\
M_{3}^{(1)}%
\end{array}
\right) =\int d{\bf v}\left(
\begin{array}{c}
1 \\
{\bf v} \\
\frac{1}{2}mV^{2}%
\end{array}
\right) \nu ({\bf r},v,t)\int d{\bf v}^{\prime }\frac{\delta g({\bf r},{\bf V%
},t\mid f)}{\delta f\left( {\bf r},{\bf v}^{\prime },t\right) }\mid
_{f=f^{(0)}}f^{(1)}\left( {\bf r},{\bf V}^{\prime },t\right) .  \label{c.6a}
\end{equation}

The terms on the right side of (\ref{c.5b}) can be identified as an
expansion in terms of polynomials of degree $2$ in the velocity. To do so,
first define a Hilbert space with scalar product
\begin{equation}
\left( a,b\right) =\int d{\bf v}\nu g^{(0)}a^{\ast }b,\hspace{0.25in}%
g^{(0)}=g\left( \mid f^{(0)}\right) .  \label{c.7}
\end{equation}
Next, define the set of functions $\left\{ \psi _{\sigma }\right\} $
\begin{equation}
\left(
\begin{array}{c}
\psi _{1} \\
{\bf \psi }_{2} \\
\psi _{3}%
\end{array}
\right) =\left(
\begin{array}{c}
\sqrt{\frac{1}{N_{1}}}1 \\
\sqrt{\frac{3}{N_{2}}}{\bf V} \\
\sqrt{\frac{1}{N_{3}}}\left( V^{2}-\frac{N_{2}}{N_{1}}\right)%
\end{array}
\right),  \label{c.8}
\end{equation}
with normalization constants
\begin{equation}
N_{1}=\left( 1,1\right) ,\hspace{0.25in}N_{2}=\left( V_{i},V_{i}\right)
=\left( 1,V^{2}\right) ,\hspace{0.25in}N_{3}=\left( \left( V^{2}-\frac{N_{2}%
}{N_{1}}\right) ,\left( V^{2}-\frac{N_{2}}{N_{1}}\right) \right) =\left(
1,V^{4}\right) -\frac{N_{2}^{2}}{N_{1}}.  \label{c.9}
\end{equation}
These functions form an orthonormal set
\begin{equation}
\left( \psi _{\sigma },\psi _{\mu }\right) =\delta _{\sigma \mu }.
\label{c.10}
\end{equation}
Equation (\ref{c.5b}) may now be written in the form
\begin{equation}
\int d{\bf v}^{\prime }\frac{\delta g({\bf r},{\bf V},t\mid f)}{\delta
f\left( {\bf r},{\bf v}^{\prime },t\right) }\mid _{f=f^{(0)}}f^{(1)}\left(
{\bf r},{\bf V}^{\prime },t\right) =g({\bf r},{\bf V},t\mid f^{(0)})\left[
e_{\sigma }\psi _{\alpha }\left( {\bf V}\right) +\frac{y(\alpha)}{nm^{2}B^{2}%
}D_{ij}({\bf V})\int d{\bf v}^{\prime }D_{ij}f^{(1)}\right].  \label{c.11}
\end{equation}
The coefficients $e_{\sigma }$ can be determined by taking the scalar
product of this equation with $\psi _{\mu }$%
\begin{equation}
e_{\mu }=\int d{\bf v}\nu \psi _{\mu }\int d{\bf v}^{\prime }\frac{\delta g(%
{\bf r},{\bf V},t\mid f)}{\delta f\left( {\bf r},{\bf v}^{\prime },t\right) }%
\mid _{f=f^{(0)}}f^{(1)}\left( {\bf r},{\bf V}^{\prime },t\right) =\int d%
{\bf v}\nu \psi _{\mu }f^{(1)}.  \label{c.12}
\end{equation}
The second equality follows from (\ref{c.6a}) and allows these terms to be
represented as a projection onto the subspace spanned by the $\left\{ \psi
_{\sigma }\right\} $%
\begin{equation}
\nu ({\bf r},V,t)\int d{\bf v}^{\prime }\frac{\delta g({\bf r},{\bf V},t\mid
f)}{\delta f\left( {\bf r},{\bf v}^{\prime },t\right) }\mid
_{f=f^{(0)}}f^{(1)}\left( {\bf r},{\bf V}^{\prime },t\right) ={\cal P}\nu
f^{(1)}+\frac{y}{nm^{2}B^{2}}\nu g^{(0)}D_{ij}\int d{\bf v}^{\prime
}D_{ij}f^{(1)},  \label{c.13}
\end{equation}
where ${\cal P}$ is a projection operator
\begin{equation}
{\cal P}X=\nu g^{(0)}\psi _{\sigma }\int d{\bf v}\psi _{\sigma }X.
\label{c.14}
\end{equation}

The linearized collision operator of (\ref{c.4}) now takes the simple form
\begin{equation}
{\cal L}f^{(1)}=\left( 1-{\cal P}\right) \nu f^{(1)}-\frac{y}{nm^{2}B^{2}}%
\nu g^{(0)}D_{ij}({\bf V})\int d{\bf v}^{\prime }D_{ij}f^{(1)}.  \label{c.15}
\end{equation}
The first term represents the fact that ${\cal L}$ has a null subspace due
to the moment conditions
\begin{equation}
\int d{\bf v}\left(
\begin{array}{c}
1 \\
{\bf V} \\
\frac{1}{2}mV^{2}%
\end{array}
\right) {\cal L}f^{(1)}=0.  \label{c.16}
\end{equation}
This is the usual BGK-like operator with a single, infintely degenerate
point in the spectrum for all functions of the orthogonal subspace. The
second term is a projection onto a specific function in the orthogonal
subspace and is the new effect of the asymmetric Gaussian approximation, or
the non-zero value of $\widetilde{B}_{ij}$.

Finally, noting that

\begin{equation}
\int d{\bf v}D_{ij}D_{ij}g^{(0)}=\frac{5}{2}nm^{2}B^{2},  \label{c.17}
\end{equation}
allows the linearized operator to be written
\begin{equation}
{\cal L}f^{(1)}=\left( 1-{\cal P}\right) \nu f^{(1)}-\frac{5}{2}y\frac{\nu
g^{(0)}D_{ij}({\bf V})\int d{\bf v}^{\prime }D_{ij}f^{(1)}}{\int d{\bf v}%
D_{ij}D_{ij}g^{(0)}}.
\end{equation}

\begin{figure}
\centerline{\epsfxsize=10cm \epsfbox{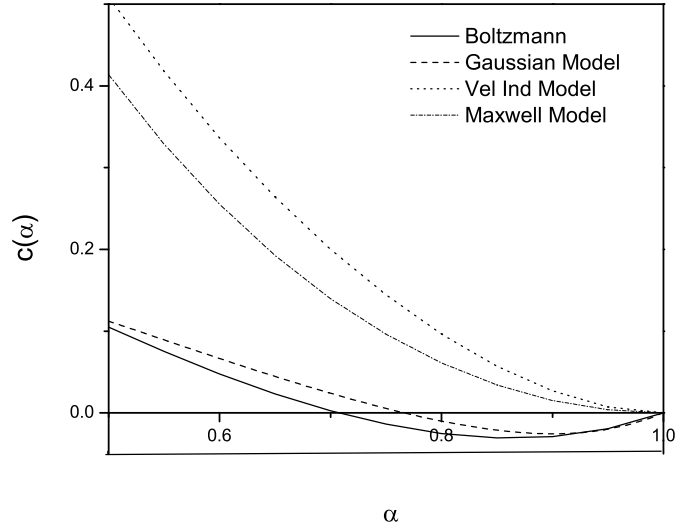}}
\caption{{\protect\small Comparison of coefficients
}$c_{G}(\protect\alpha)${\protect\small,}
$c_{G1}(\protect\alpha)$ {\protect\small and
}$c_{M}(\protect\alpha)$ {\protect\small with }$c_{B}(
\protect\alpha).$}
\end{figure}

\begin{figure}
\centerline{\epsfxsize=10cm \epsfbox{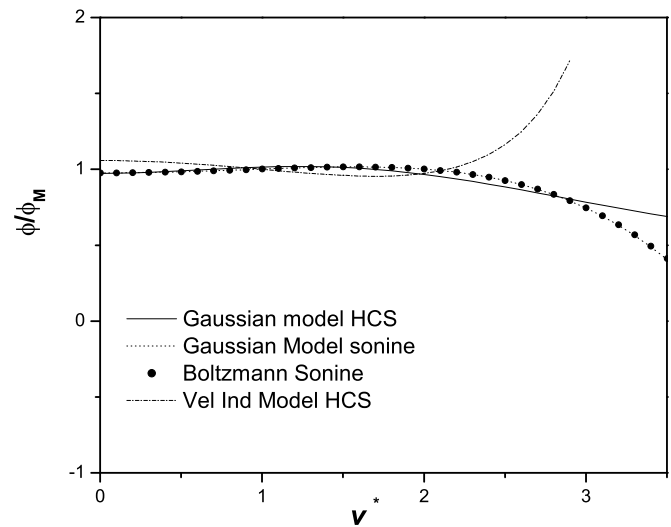}}
\caption{{\protect\small Comparison of the HCS divided by the
Maxwellian for the Gaussian model and the velocity independent
collision frequency model with the Sonine approximation using both
}$c_{G}(\protect\alpha)$ {\protect\small and }
$c_{B}(\protect\alpha)$ {\protect\small for } $\protect\alpha$
{\protect\small = 0.8.}}
\end{figure}

\begin{figure}
\centerline{\epsfxsize=10cm \epsfbox{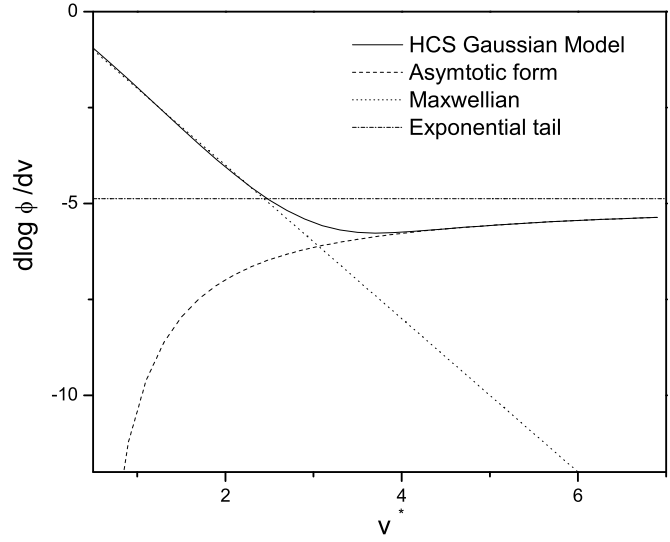}}
\caption{{\protect\small Illustration of the cross over of the HCS
for the Gaussian model to the intermediate form, eq(\ref{4.31}),
by plotting }$\partial _{v^{\ast }}\ln \phi (v^{\ast })$ {\small
 for the HCS, the asymptotic form and the Maxwellian for }$\alpha
=0.8$.}
\end{figure}


\begin{figure}
\centerline{\epsfxsize=10cm \epsfbox{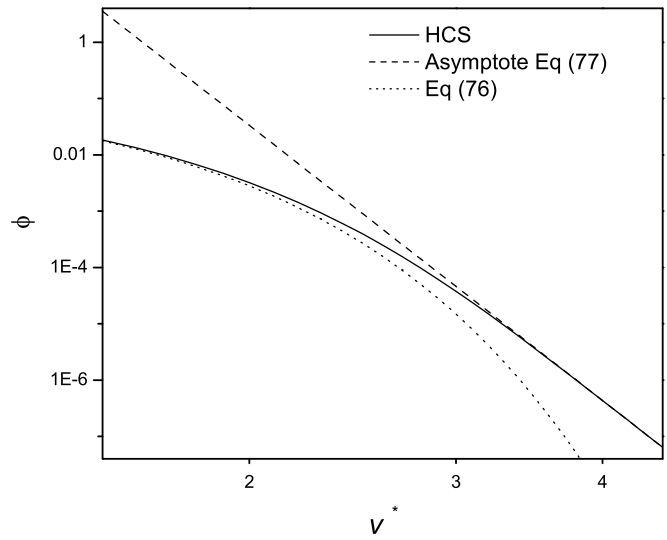}}
\caption{{\protect\small Illustration of crossover of the HCS for
the velocity independent model from the Gaussian given by
Eq.(\ref{4.21b}) to an algebraic decay, Eq.(\ref{4.22}) for
}$\protect\alpha $ = 0.8.}
\end{figure}

\begin{figure}
\centerline{\epsfxsize=10cm \epsfbox{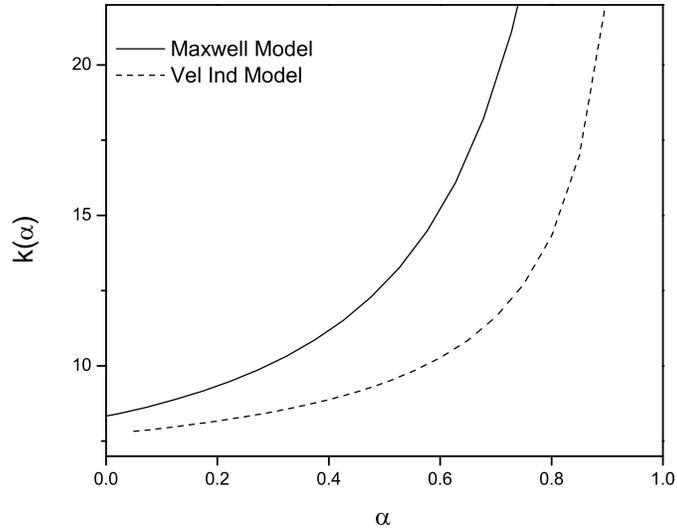}}
\caption{{\protect\small The exponent of algebraic decay for the
Maxwell model, Eq.(\ref{2.18}), and velocity independent Gaussian
model, Eq.(\ref{4.22}) plotted as a function of }$\protect\alpha
$.}
\end{figure}




\begin{figure}
\centerline{\epsfxsize=10cm \epsfbox{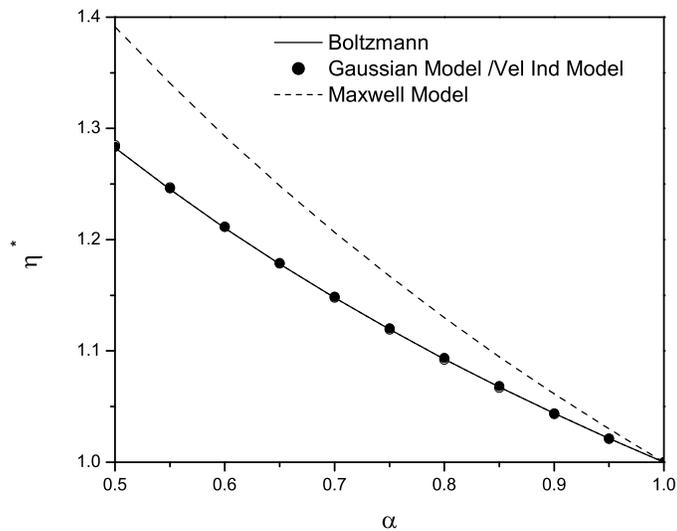}}
\caption{{\protect\small Plot of the fitted dimensionless viscosity} $%
\protect\eta^{\ast}=\protect\eta/\protect\eta_{0}$ {\protect\small %
for the velocity dependent and velocity independent collision
frequency Gaussian models with the Boltzmann and Maxwell model
results.}}
\end{figure}

\begin{figure}
\centerline{\epsfxsize=10cm \epsfbox{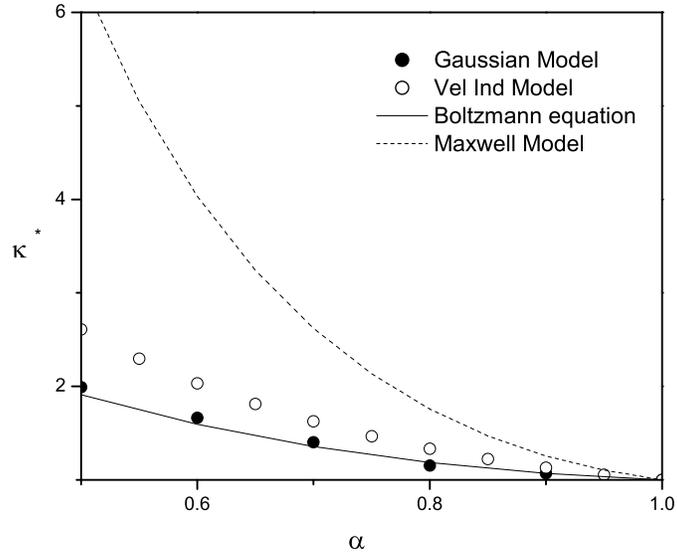}}
\caption{{\protect\small Comparison of dimensionless thermal conductivity} $%
\protect\kappa^{\ast}= \protect\kappa/\protect\kappa_{0}$
{\protect\small as calculated from the velocity dependent and
velocity independent collision frequency Gaussian models with the
Boltzmann  and Maxwell model results.}}
\end{figure}

\begin{figure}
\centerline{\epsfxsize=10cm \epsfbox{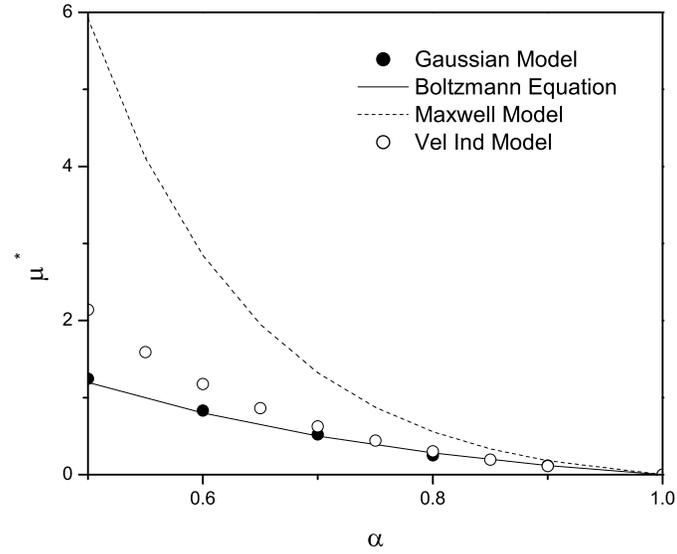}}
\caption{{\protect\small Comparison of }$\protect\mu^{\ast}=\protect%
\mu n/T\protect\kappa_{0}$ {\protect\small as calculated from
the Gaussian models with the Boltzmann and Maxwell model
results.}}
\end{figure}

\begin{figure}
\centerline{\epsfxsize=10cm \epsfbox{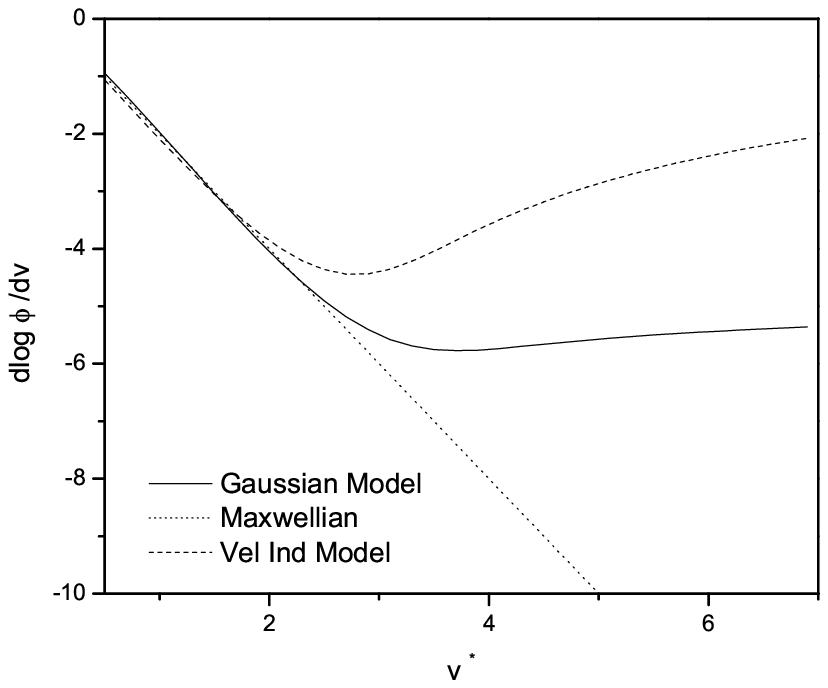}}
\caption{{\protect\small Plot of }$\partial _{v^{\ast }}\ln \phi
(v^{\ast })$ {\protect\small for the Gaussian model and velocity
independent Gaussian model and the Maxwellian for }$\alpha=0.8$.}
\end{figure}



\begin{references}
\bibitem{hist2} For the case of a velocity independent collision frequency
and elastic collisions the Gaussian model reduces to the ES model of
reference $\cite{ES66}$. For inelastic collisions it constitutes a
generalization of reference $\cite{BMyD96}$ to the non-spherical Gaussian
model.
\end{references}
\end{document}